\documentclass[12pt]{article}
\usepackage{epsfig}
\begin{document}
%%%%%%%%%%%%%%%%%%%%%%%%%%%%%%%%%%%%%%%%%%%%%%%%%%%%%%%%%%%%%%%%%%%%%%%%%
\title{Probing the hidden-bottom pentaquark resonances in photonuclear bottomonium
production near threshold: differential observables}
\author{E. Ya. Paryev \\
{\it Institute for Nuclear Research of the Russian Academy of Sciences}\\
{\it Moscow, Russia}}
%==============================================================
%%==============================================================

\renewcommand{\today}{}
\maketitle

\begin{abstract}
We study the near-threshold $\Upsilon(1S)$ meson photoproduction from protons and nuclei
by considering incoherent direct non-resonant (${\gamma}p \to {\Upsilon(1S)}p$,
${\gamma}n \to {\Upsilon(1S)}n$) and two-step resonant
(${\gamma}p \to P^+_{bi} \to {\Upsilon(1S)}p$, ${\gamma}n \to P^0_{bi} \to {\Upsilon(1S)}n$,
$i=1$, 2, 3; $P^{+,0}_{b1}=P^{+,0}_{b}(11080)$,
$P^{+,0}_{b2}=P^{+,0}_{b}(11125)$, $P^{+,0}_{b3}=P^{+,0}_{b}(11130)$) bottomonium production processes
with the main goal of clarifying the possibility to observe the non-strange hidden-bottom pentaquark
states $P^{+,0}_{bi}$ in this production via differential observables.
We calculate the absolute excitation functions, energy and momentum distributions
for the non-resonant, resonant and for the combined (non-resonant plus resonant) production
of $\Upsilon(1S)$ mesons on protons, on carbon and tungsten target nuclei at near-threshold incident
photon beam energies by assuming the spin-parity assignments of the hypothetical hidden-bottom resonances
$P^{+,0}_{b}(11080)$, $P^{+,0}_{b}(11125)$ and $P^{+,0}_{b}(11130)$ as $J^P=(1/2)^-$, $J^P=(1/2)^-$
and $J^P=(3/2)^-$ within four different realistic choices for the branching ratios
of their decays to the ${\Upsilon}(1S)p$ and ${\Upsilon}(1S)n$ modes (0.125, 0.25, 0.5 and 1\%)
as well as for two options for the background contribution. We demonstrate that the measurements of
these combined observables on proton and nuclear targets in the near-threshold energy region in future
experiments at the planned high-luminosity electron-ion colliders EIC and EicC in the US and China should
provide evidence for the existence of the above hidden-bottom pentaquark resonances as well as clarify
their decay rates.
\end{abstract}

\newpage

\section*{1. Introduction}

The discovery of $X(3872)$ resonance (also known as $\chi_{c1}(3872)$ [1]) by the Belle Collaboration in
2003 [2] as a narrow peak in the vicinity of the $D^0{\bar D}^{*0}$ mass threshold in the ${J/\psi}{\pi^+}\pi^-$
invariant mass distribution in exclusive $B^{\pm} \to K^{\pm}X(3872) \to K^{\pm}({J/\psi}{\pi^+}\pi^-)$ decays
\footnote{$^)$It was also confirmed in many other high-energy experiments.}$^)$
has opened a new era for the study of exotic heavy hadrons, which exhibit properties incompatible with the
predictions of the traditional quark model for quark-antiquark mesons and three-quark baryons.
They are composed of four or five quarks (and antiquarks) and usually are named as tetraquark and pentaquark
states, respectively. Since then, many unconventional charmonium- and bottomonium-like states, the so-called
$X, Y, Z$ mesons
\footnote{$^)$The $X$ and $Z$ correspond to tetraquark ground states in $S$-wave (positive parity),
the $Y$ to negative parity $P$-waves [3].}$^)$
as well as the pentaquark $P_c$ states have been observed. For a comprehensive presentation of the status of
exotic hadron physics, we refer the reader to Refs. [3--8] for a review. The most recently observed
exotic states in the charm sector are: the hidden-charm non-strange pentaquark resonances $P^+_{c}(4312)$,
$P^+_{c}(4440)$, $P^+_{c}(4457)$ [9] and $P^+_{c}(4337)$ [10] with minimal quark structure of $|P^+_c>=|uudc{\bar c}>$
in the ${J/\psi}p$ mass spectrum; the hidden-charm strange pentaquark states $P^0_{cs}(4459)$ [11] and $P^0_{cs}(4338)$ [12] with minimal quark content $|P^0_{cs}>=|udsc{\bar c}>$ in the ${J/\psi}\Lambda$ decay mode;
the fully charmed tetraquarks $X(6900)$ [13--17] and $X(6600)$, $X(7300)$ [14, 15]
(or $T_{cc{\bar c}{\bar c}}$) with minimum quark content $|cc{\bar c}{\bar c}>$ in the di-$J/\psi$ channel;
the new charm-strange spin-0 and spin-1 resonances $X_0(2900)$ and $X_1(2900)$ in the $D^-K^+$ channel with minimum valence quark contents of four different flavors, i.e. $|X_{0,1}(2900)>=|ud{\bar s}{\bar c}>$ [18, 19]; the firstly observed narrow doubly-charmed tetraquark $T^+_{cc}(3875)$ with the typical quark configuration $|T^+_{cc}(3875)>=|cc{\bar u}{\bar d}>$ in the $D^0D^0\pi^+$ invariant mass spectrum [20, 21];
the doubly-charged open-charm tetraquark $T^{\alpha}_{c{\bar s}0}(2900)^{++}$ and its neutral isospin partner $T^{\alpha}_{c{\bar s}0}(2900)^{0}$
in the $D^+_{s}{\pi^+}$ and $D^+_{s}{\pi^-}$ channels with minimal quark constituents $|c{\bar s}u{\bar d}>$ and
$|c{\bar s}{\bar u}d>$, respectively, [22, 23]; the strange charged hidden-charm tetraquark state with open strangeness $Z_{cs}(3985)^-$ [24] and its neutral isospin partner $Z_{cs}(3985)^0$ [25] with  minimal quark configurations $|Z_{cs}(3985)^->=|{\bar u}sc{\bar c}>$ and $|Z_{cs}(3985)^0>=|{\bar d}sc{\bar c}>$, respectively;
the exotic states $Z_{cs}(4000)^+$ and $Z_{cs}(4220)^+$ with a new quark composition $|u{\bar s}c{\bar c}>$ in the
${J/\psi}K^+$ channel [26]. Since the experimental discovery of the $X(3872)$, the internal structure of the exotic
hadronic resonances (loosely bound hadron molecules or compact multiquark states) has been intensely debated.
However, a compelling understanding of the nature of these resonances is still lacking. Thus,
in a molecular scenario, due to the closeness of the observed
$P^+_{c}(4312)$ and $P^+_c(4440)$, $P^+_c(4457)$ masses to the ${\Sigma^+_c}{\bar D}^0$ and
${\Sigma^+_c}{\bar D}^{*0}$ thresholds, the $P^+_c(4312)$
resonance can be, in particular, interpreted as an S-wave ${\Sigma^+_c}{\bar D}^0$ bound state, while the
$P^+_c(4440)$ and $P^+_c(4457)$ as S-wave ${\Sigma^+_c}{\bar D}^{*0}$ bound molecular states [27--33], though
there are alternative explanations too [7, 8, 34--36].

Meanwhile, several other types of tetraquarks and pentaquarks have been theoretically studied recently, such as
bottomonium-like [37], doubly-heavy [38] and doubly-bottom [39, 40] tetraquarks as well as the hidden-strange [41] and the hidden-bottom [42--47] pentaquarks, the hidden-bottom pentaquarks with strangeness [42, 48],
the hidden-charm pentaquark states with double and triple strangeness [35, 49--52]; the single [53] and
doubly-charmed [54--56] pentaquarks; the fully-charm and -bottom pentaquarks [57]; charmed-bottom pentaquarks [58]
and pentaquark states with open charm and bottom flavors [59]. The observation of these exotic candidates
and, in particular, the hidden-bottom pentaquarks $P_b$ as the bottom partners of the experimentally observed hidden-charm pentaquarks $P_c$ in the charm sector is of great importance to derive a complete and unified picture of all flavors and, hence, to gain a deeper insight into the mechanism of low-energy QCD.

In a molecular picture, the structure of the $P_b$ states can be predicted, starting from that of the $P_c$ resonances,
by replacing the $c{\bar c}$ pair on the bottom-antibottom $b{\bar b}$ pair as well as
the non-strange $D(D^*)$ mesons on $B(B^*)$ ones and
the charmed baryons by the bottom ones. Based on the classification of hidden-charm pentaquarks
composed by a single charm baryon and $D(D^*)$ mesons, such prediction has been performed in Ref. [42]
using the hadronic molecular approach. As a result, the classification of hidden-bottom pentaquarks
composed by a single bottom baryon and $B(B^*)$ mesons has been presented here. Accordingly,
the charged hidden-bottom partners $P^+_b(11080)$, $P^+_b(11125)$ and $P^+_b(11130)$ of the observed
hidden-charm pentaquarks $P^+_c(4312)$, $P^+_c(4440)$ and $P^+_c(4457)$, having the quark structure
$|P^+_b>=|uudb{\bar b}>$, were predicted to exist, with masses of 11080, 11125 and 11130 MeV, respectively.
Moreover, the predictions for the neutral hidden-bottom counterparts $P^0_b(11080)$, $P^0_b(11125)$ and
$P^0_b(11130)$ of the unobserved hidden-charm states $P^0_c(4312)$, $P^0_c(4440)$ and $P^0_c(4457)$ with
the quark structure $|P^0_b>=|uddb{\bar b}>$ were provided in [42] as well.
The masses of these new exotic heavy pentaquarks are all above 11 GeV, so they may be observed in the
${\Upsilon(1S)}p$ and ${\Upsilon(1S)}n$ final states. Due to very large mass of the hidden-bottom pentaquarks, it is relatively difficult to search for them experimentally compared with the $P_c$ states. Unlike these charm states,
the bottom states can hardly be produced via the weak decays of more heavier hadrons, because of the very rare
events of such hadrons [60]. Therefore, they can only be produced in high-energy hadron-hadron, electron-hadron
and photon-hadron collisions. Thus, in Ref. [61] pion and photon-induced productions of hidden-bottom
pentaquarks on protons were studied, the calculations suggest that it would be possible to look for these states
at J-PARC and EicC facilities. The exploration of searching for a typical $P_b$ state in the ${\gamma}p \to {\Upsilon(1S)}p$ reaction has been carried out in Ref. [62]. This shows a promising potential to observe it at the electron-ion colliders. In our work [63], we have calculated the absolute excitation functions (the total cross sections) for the non-resonant, resonant and for the combined (non-resonant plus resonant) photoproduction of $\Upsilon(1S)$ mesons off protons as well as off carbon and lead target nuclei at near-threshold incident photon energies
within five different scenarios for the branching ratios of the decays $P^{+,0}_{b}(11080/11125/11130) \to {\Upsilon(1S)}p(n)$, namely: 1, 2, 3, 5, 10\%. It was found that the resonant $\Upsilon(1S)$ production cross
sections would be much larger than the non-resonant ones only if these branching ratios $\sim$ 2\% and higher.
Otherwise, the role of the pentaquarks $P^{+,0}_{b}(11080/11125/11130)$ considered in bottomonium photoproduction total cross sections on protons and nuclei is insignificant and they cannot be observed through their detailed scan
in the near-threshold energy region.

To get a robust enough information for or against the existence of the hidden-bottom pentaquarks and to understand their better, it is crucial to investigate the possibility of their observation by measuring not only the excitation functions for $\Upsilon(1S)$ mesons from photon-induced reactions on protons and nuclei at near-threshold photon energies, predicted in Ref. [63], but also their differential observables -- the energy and momentum distributions in these reactions, not predicted in the previous paper [63]. The predictions of the latter ones are the main purpose of the present work. It is mainly based on the models, developed in Refs. [63, 64].
These predictions may serve as guidance for future $\Upsilon(1S)$ photoproduction experiments at the proposed
electron-ion colliders EIC and EicC in the U.S. and China.

\section*{2. Model}

\subsection*{2.1. Direct non-resonant $\Upsilon(1S)$ production processes}

Direct non-resonant bottomonium photoproduction on nuclear targets in the near-threshold laboratory incident
photon beam energy region $57.2~{\rm GeV} \le E_{\gamma} \le 68.8~{\rm GeV}$
\footnote{$^)$Which corresponds to the center-of-mass energy W of the photon-proton system
$10.4~{\rm GeV} \le W \le 11.4~{\rm GeV}$
and in which the hidden-bottom pentaquarks $P_{b}^{+}(11080)$, $P_{b}^{+}(11125)$ and
$P_{b}^{+}(11130)$ are concentrated and where they can be observed in the ${\gamma}p$ and ${\gamma}A$ reactions
at the proposed electron-ion colliders in China (EicC) [65, 66] and US (EIC) [67--70].
We remind that the threshold (resonant) energies $E_{\gamma}^{\rm R1}$,
$E^{\rm R2}_{\gamma}$ and $E^{\rm R3}_{\gamma}$ for the production of the $P_{b}^{+}(11080)$,
$P_{b}^{+}(11125)$ and $P_{b}^{+}(11130)$ resonances with central values of their masses
$M_{b1}=11080$ MeV, $M_{b2}=11125$ MeV and $M_{b3}=11130$ MeV
on a free target proton being at rest are $E^{\rm R1}_{\gamma}=64.952$ GeV, $E^{\rm R2}_{\gamma}=65.484$ GeV
and $E^{\rm R3}_{\gamma}=65.544$ GeV, respectively.}$^)$
may proceed via the following elementary processes with the lowest free production
threshold ($\approx$57.2 GeV) [63]:
%formula(1)
\begin{equation}
\gamma+p \to \Upsilon(1S)+p,
\end{equation}
%formula(2)
\begin{equation}
\gamma+n \to \Upsilon(1S)+n.
\end{equation}
At first, we consider the kinematic characteristics, for example, of $\Upsilon(1S)$ mesons and final protons -- laboratory polar production angles and momenta (total energies), allowed in the process (1) in the simpler case of a
free target proton being at rest at an incident resonant photon energy of 65.484 GeV, which is needed for the
production of the $P_{b}^+(11125)$ resonance with central value of its mass $M_{b2}=11125$ MeV in this process. The kinematics of two-body reaction with a threshold
(as in our present case) indicate that the laboratory polar $\Upsilon(1S)$ and final proton production angles $\theta_{\Upsilon(1S)}$ and $\theta_p$
vary from 0 to a maximal values $\theta^{\rm max}_{\Upsilon(1S)}$ and $\theta^{\rm max}_{p}$, respectively, i.e.:
%formula(3)
\begin{equation}
     0 \le \theta_{\Upsilon(1S)} \le \theta^{\rm max}_{\Upsilon(1S)},
\end{equation}
%formula(4)
\begin{equation}
     0 \le \theta_{p} \le \theta^{\rm max}_{p},
\end{equation}
where
%formula(5)
\begin{equation}
 \theta^{\rm max}_{\Upsilon(1S)}={\rm arcsin}[(\sqrt{s}p^{*}_{\Upsilon(1S)})/(m_{\Upsilon(1S)}p_{\gamma})],
\end{equation}
%formula(6)
\begin{equation}
 \theta^{\rm max}_{p}={\rm arcsin}[(\sqrt{s}p^{*}_{p})/(m_{p}p_{\gamma})].
\end{equation}
Here, the $\Upsilon(1S)$ and proton c.m. momenta $p^*_{\Upsilon(1S)}$ and $p^{*}_{p}$ are determined by the equation
%FORMULA(7)
\begin{equation}
p_{\Upsilon(1S)}^*=p^{*}_{p}=\frac{1}{2\sqrt{s}}\lambda(s,m_{\Upsilon(1S)}^{2},m_{p}^2),
\end{equation}
in which the quantities $m_{\Upsilon(1S)}$ and $m_{p}$ are the free space bottomonium and
proton masses and the vacuum collision energy squared $s$ and function $\lambda(x,y,z)$
are defined, respectively, by the formulas
%formula(8)
\begin{equation}
  s=W^2=(E_{\gamma}+m_p)^2-{\bf p}_{\gamma}^2=m_p^2+2m_pE_{\gamma},
\end{equation}
%FORMULA (9)
\begin{equation}
\lambda(x,y,z)=\sqrt{{\left[x-({\sqrt{y}}+{\sqrt{z}})^2\right]}{\left[x-
({\sqrt{y}}-{\sqrt{z}})^2\right]}}.
\end{equation}
The quantity ${\bf p}_{\gamma}$ in Eq. (8) denotes the momentum of the incident photon
in the laboratory system ($E_{\gamma}=|{\bf p}_{\gamma}|=p_{\gamma}$).
From Eqs. (5), (6) we get that
%formula(10)
\begin{equation}
\theta^{\rm max}_{\Upsilon(1S)}=1.308^{\circ}\,\,\,\,{\rm and}\,\,\,\
\theta^{\rm max}_{p}=13.307^{\circ}
\end{equation}
at incident resonant photon beam energy of 65.484 GeV.
%%%%%%%%%%%%%%%%%%%%%%%%%%%%%%%%%%%%%%%%%%%%%%%%%%%%%%%%%%%
\begin{figure}[htb]
\begin{center}
\includegraphics[width=16.0cm]{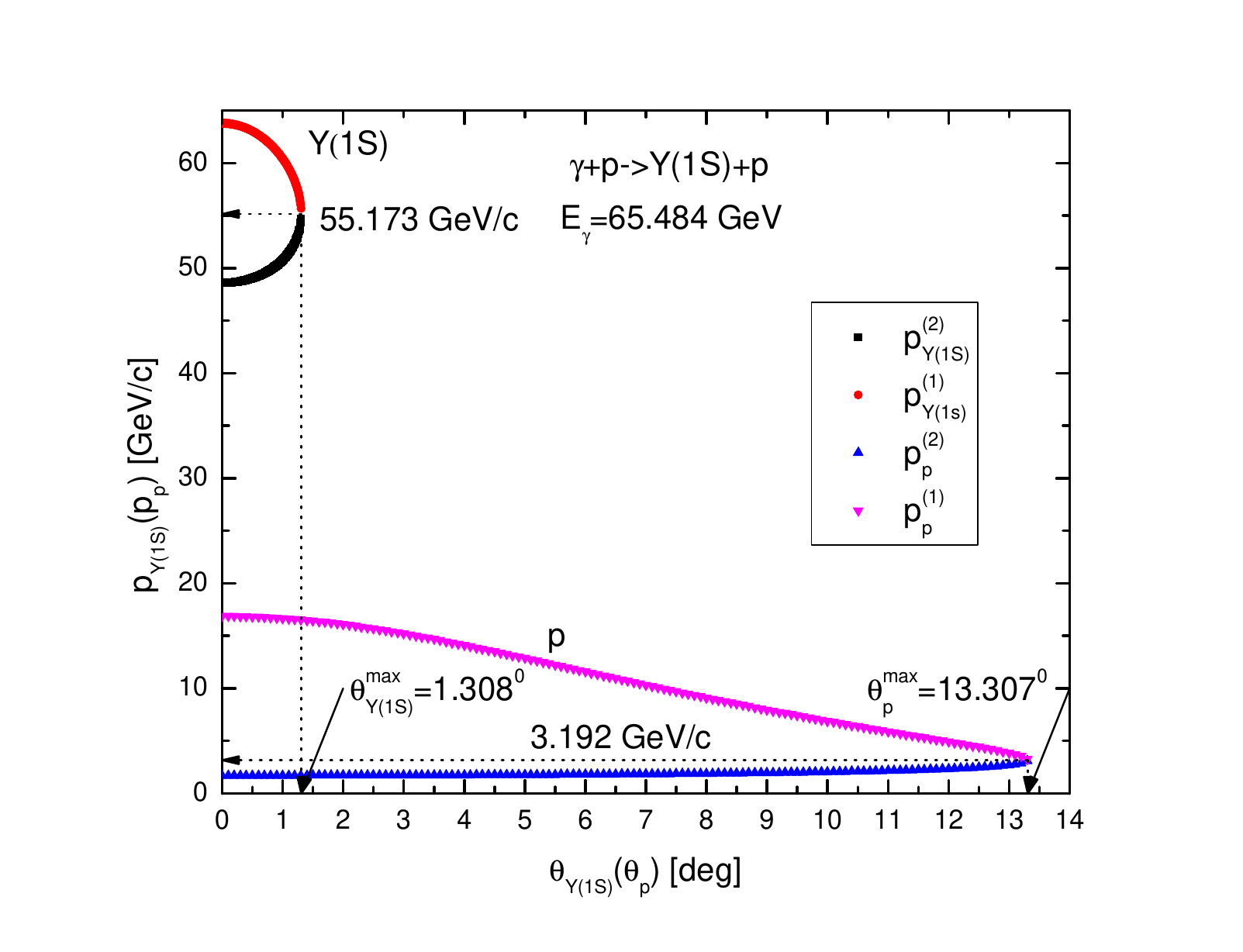}
\vspace*{-2mm} \caption{(Color online.) Plot of the allowed final $\Upsilon(1S)$ meson and proton momenta
in the direct non-resonant ${\gamma}p \to {\Upsilon(1S}p$ reaction, occuring
in the laboratory system in the free space at initial photon energy of 65.484 GeV,
as functions of their production angles with respect to the photon beam direction in this system.}
\label{void}
\end{center}
\end{figure}
%%%%%%%%%%%%%%%%%%%%%%%%%%%%%%%%%%%%%%%%%%%%%%%%%%%%%%%%%%%
Imposing energy-momentum conservation in
the free space reaction (1), one can find two different solutions for the laboratory
$\Upsilon(1S)$ meson and final proton momenta $p_{\Upsilon(1S)}$ and $p_{p}$ at given laboratory
polar production angles
$\theta_{\Upsilon(1S)}$ and $\theta_{p}$, belonging, correspondingly, to the angular intervals (3) and (4):
%formula(11)
\begin{equation}
p^{(1,2)}_{\Upsilon(1S)}(\theta_{\Upsilon(1S)})=
\frac{p_{\gamma}\sqrt{s}E^{*}_{\Upsilon(1S)}\cos{\theta_{\Upsilon(1S)}}\pm
(E_{\gamma}+m_p)\sqrt{s}\sqrt{p^{*2}_{\Upsilon(1S)}-{\gamma^2_{\rm cm}}{v^2_{\rm
cm}}m^2_{\Upsilon(1S)}\sin^2{\theta_{\Upsilon(1S)}}}}{(E_{\gamma}+m_p)^2-p^2_{\gamma}\cos^2{\theta_{\Upsilon(1S)}}},
\end{equation}
%formula(12)
\begin{equation}
p^{(1,2)}_{p}(\theta_{p})=
\frac{p_{\gamma}\sqrt{s}E^{*}_{p}\cos{\theta_{p}}\pm
(E_{\gamma}+m_p)\sqrt{s}\sqrt{p^{*2}_{p}-{\gamma^2_{\rm cm}}{v^2_{\rm
cm}}m^2_{p}\sin^2{\theta_{p}}}}{(E_{\gamma}+m_p)^2-p^2_{\gamma}\cos^2{\theta_{p}}}.
\end{equation}
Here, ${\gamma_{\rm cm}}=(E_{\gamma}+m_p)/\sqrt{s}$, $v_{\rm cm}=p_{\gamma}/(E_{\gamma}+m_p)$,
the $\Upsilon(1S)$ and final proton total c.m. energies $E^{*}_{\Upsilon(1S)}$ and $E^{*}_{p}$ are defined as:
$E^{*}_{\Upsilon(1S)}=\sqrt{m^2_{\Upsilon(1S)}+p^{*2}_{\Upsilon(1S)}}$ and $E^{*}_{p}=\sqrt{m^2_{p}+p^{*2}_{p}}$, respectively. Sign "+" in the numerators of Eqs. (11), (12)
corresponds to the first solutions $p^{(1)}_{\Upsilon(1S)}$, $p^{(1)}_{p}$
and sign "-" - to the second ones $p^{(2)}_{\Upsilon(1S)}$, $p^{(2)}_{p}$.
Inspecting the expressions (11) and (12), we come to the conclusion that the first solutions
$p^{(1)}_{\Upsilon(1S)}$, $p^{(1)}_{p}$ as well as the second ones $p^{(2)}_{\Upsilon(1S)}$, $p^{(2)}_{p}$
have different dependencies, respectively, on the production angles $\theta_{\Upsilon(1S)}$ and $\theta_{p}$
within the angular intervals [0, $\theta_{\Upsilon(1S)}^{\rm max}]$ and [0, $\theta_{p}^{\rm max}]$.
Namely, the former ones drop and the latter increase as the production angles
$\theta_{\Upsilon(1S)}$ and $\theta_{p}$ increase in these intervals (cf. Fig. 1) and
%formula(13)
\begin{equation}
p^{(1)}_{\Upsilon(1S)}(\theta_{\Upsilon(1S)}^{\rm max})=p^{(2)}_{\Upsilon(1S)}(\theta_{\Upsilon(1S)}^{\rm max})=
p_{\Upsilon(1S)}(\theta_{\Upsilon(1S)}^{\rm max}),
\end{equation}
%formula(14)
\begin{equation}
p^{(1)}_{p}(\theta_{p}^{\rm max})=p^{(2)}_{p}(\theta_{p}^{\rm max})=
p_{p}(\theta_{p}^{\rm max}),
\end{equation}
 where
%formula(15)
\begin{equation}
p_{\Upsilon(1S)}(\theta_{\Upsilon(1S)}^{\rm max})=\sqrt{E_{\Upsilon(1S)}^2(\theta_{\Upsilon(1S)}^{\rm max})-m_{\Upsilon(1S)}^2},\,\,
E_{\Upsilon(1S)}(\theta_{\Upsilon(1S)}^{\rm max})={\gamma_{\rm cm}}m_{\Upsilon(1S)}^2/E_{\Upsilon(1S)}^*;
\end{equation}
%formula(16)
\begin{equation}
p_{p}(\theta_{p}^{\rm max})=\sqrt{E_{p}^2(\theta_{p}^{\rm max})-m_{p}^2},\,\,
E_{p}(\theta_{p}^{\rm max})={\gamma_{\rm cm}}m_{p}^2/E_{p}^*.
\end{equation}
According to Eqs. (15), (16), for $E_{\gamma}=65.484$ GeV we get then that
$p_{\Upsilon(1S)}(\theta_{\Upsilon(1S)}^{\rm max})=55.173$ GeV/c and $p_{p}(\theta_{p}^{\rm max})=3.192$ GeV/c
(cf. Fig. 1).
This figure shows that the kinematically allowed bottomonium laboratory momentum $p_{\Upsilon(1S)}$
and total energy $E_{\Upsilon(1S)}$ ($E_{\Upsilon(1S)}=\sqrt{m^2_{\Upsilon(1S)}+p^{2}_{\Upsilon(1S)}})$
in the direct process (1), proceeding on the vacuum target proton at rest, vary within the
following momentum and energy ranges at given initial photon energy:
%formula(17)
\begin{equation}
p^{(2)}_{\Upsilon(1S)}(0^{\circ}) \le p_{\Upsilon(1S)} \le p^{(1)}_{\Upsilon(1S)}(0^{\circ}),
\end{equation}
%formula(18)
\begin{equation}
E^{(2)}_{\Upsilon(1S)}(0^{\circ}) \le E_{\Upsilon(1S)} \le E^{(1)}_{\Upsilon(1S)}(0^{\circ}),
\end{equation}
where the quantities $p^{(1,2)}_{\Upsilon(1S)}(0^{\circ})$ and $E^{(1,2)}_{\Upsilon(1S)}(0^{\circ})$ are
the $\Upsilon(1S)$ momenta and energies at zero polar $\Upsilon(1S)$ production angle in the laboratory frame.
From Eq. (11), we obtain:
%formula(19)
\begin{equation}
p^{(1,2)}_{\Upsilon(1S)}(0^{\circ})=
{\gamma_{\rm cm}}E^{*}_{\Upsilon(1S)}(v_{\rm cm}\pm v^{*}_{\Upsilon(1S)}),
\end{equation}
%formula(20)
\begin{equation}
E^{(1,2)}_{\Upsilon(1S)}(0^{\circ})=
{\gamma_{\rm cm}}(E^{*}_{\Upsilon(1S)}\pm v_{\rm cm}p^{*}_{\Upsilon(1S)}).
\end{equation}
Here,
%formula(21)
\begin{equation}
v^{*}_{\Upsilon(1S)}=p^{*}_{\Upsilon(1S)}/E^{*}_{\Upsilon(1S)}.
\end{equation}
In analogy to $\Upsilon(1S)$ production, for final proton, the allowed momentum and energy coverages look
like those of (17)--(21), but in which one needs to make the substitution $\Upsilon(1S) \to p$.
In line with this fact and Eqs. (17)--(21), for $E_{\gamma}=65.484$ GeV we have:
%formula(22)
\begin{equation}
48.594~\le p_{\Upsilon(1S)}~\le 63.774~{\rm GeV/c},
\end{equation}
%formula(23)
\begin{equation}
49.507~\le E_{\Upsilon(1S)}~\le 64.472~{\rm GeV};
\end{equation}
%formula(24)
\begin{equation}
1.710~\le p_{p}~\le 16.890~{\rm GeV/c},
\end{equation}
%formula(25)
\begin{equation}
1.951~\le E_{p}~\le 16.916~{\rm GeV}.
\end{equation}
Since the neutron mass is approximately equal to the proton mass, the kinematical characteristics of
the $\Upsilon(1S)$ mesons and final neutrons, produced in the reaction (2), are close to those of
final particles ($\Upsilon(1S)$ mesons and protons) in the process (1).
Evidently, the binding of target nucleons and their Fermi motion will distort the distributions of the
outgoing high-momentum (and high-energy) $\Upsilon(1S)$ mesons and nucleons
as well as lead to a wider accessible momentum and energy intervals compared to those given
above by Eqs. (17)--(25). With this, we will neglect the modification of the final high-momentum
$\Upsilon(1S)$ mesons (cf. [71--74]) and outgoing nucleons [75] in the nuclear medium in the case when
the reactions (1), (2) proceed on a nucleons embedded in a nuclear target.

Disregarding the absorption of incident photons in the energy range of interest as well as the $\Upsilon(1S)$
meson quasielastic rescatterings on intranuclear nucleons, accounting for the fact that the $\Upsilon(1S)$
mesons are produced at small laboratory polar angles (see above) and describing the bottomonium final-state absorptions in the nuclear matter by the effective cross section $\sigma_{{\Upsilon(1S)}N}$
\footnote{$^)$For this cross section we have used the value $\sigma_{{\Upsilon(1S)}N}=1$ mb motivated by the results
from Ref. [76], according to which the $\Upsilon(1S)$--nucleon total cross section in the bottomonium momentum range
considered here (cf. Eq. (22)) is expected to be of the order of 0.5--1.5 mb for different models for the dipole cross section.}$^)$,
we can represent the inclusive differential cross section for the production of $\Upsilon(1S)$ mesons with the vacuum
momentum ${\bf p}_{\Upsilon(1S)}$ off nuclei in the direct non-resonant photon-induced reaction channels (1), (2) as follows (cf. Ref. [64]):
%formula(26)
\begin{equation}
\frac{d\sigma_{{\gamma}A\to {\Upsilon(1S)}X}^{({\rm dir})}({\bf p}_{\gamma},{\bf p}_{\Upsilon(1S)})}
{d{\bf p}_{\Upsilon(1S)}}=I_{V}[A,\sigma_{{\Upsilon(1S)}N}]
\left<\frac{d\sigma_{{\gamma}p \to {\Upsilon(1S)}p}({\bf p}_{\gamma},{\bf p}_{\Upsilon(1S)})}{d{\bf p}_{\Upsilon(1S)}}\right>_A,
\end{equation}
where
%formula(27)
\begin{equation}
I_{V}[A,\sigma]=2{\pi}A\int\limits_{0}^{R}r_{\bot}dr_{\bot}
\int\limits_{-\sqrt{R^2-r_{\bot}^2}}^{\sqrt{R^2-r_{\bot}^2}}dz
\rho(\sqrt{r_{\bot}^2+z^2})
\exp{\left[-A{\sigma}\int\limits_{z}^{\sqrt{R^2-r_{\bot}^2}}\rho(\sqrt{r_{\bot}^2+x^2})dx\right]},
\end{equation}
%formula(28)
\begin{equation}
\left<\frac{d\sigma_{{\gamma}p \to {\Upsilon(1S)}p}({\bf p}_{\gamma},{\bf p}_{\Upsilon(1S)})}{d{\bf p}_{\Upsilon(1S)}}\right>_A=
\int\int
P_A({\bf p}_t,E)d{\bf p}_tdE
\left[\frac{d\sigma_{{\gamma}p \to {\Upsilon(1S)}p}(\sqrt{s^*},{\bf p}_{\Upsilon(1S)})}{d{\bf p}_{\Upsilon(1S)}}\right]
\end{equation}
and
%formula(29)
\begin{equation}
  s^*=(E_{\gamma}+E_t)^2-({\bf p}_{\gamma}+{\bf p}_t)^2,
\end{equation}
%formula(30)
\begin{equation}
   E_t=M_A-\sqrt{(-{\bf p}_t)^2+(M_{A}-m_{p}+E)^{2}}.
\end{equation}
Here, $d\sigma_{{\gamma}p\to {\Upsilon(1S)}p}(\sqrt{s^*},{\bf p}_{\Upsilon(1S)})/d{\bf p}_{\Upsilon(1S)}$
is the off-shell differential cross section for the production of $\Upsilon(1S)$ meson in process (1)
at the "in-medium" ${\gamma}p$ c.m. energy $\sqrt{s^*}$
\footnote{$^)$In Eq. (26), we assume that the cross sections for $\Upsilon(1S)$ meson production in
${\gamma}p$ and ${\gamma}n$ interactions are equal to each other [63].}$^)$;
$\rho({\bf r})$ and $P_A({\bf p}_t,E)$ are normalized to unity the nucleon density and the nuclear
spectral function (they are given in Refs. [77, 78]) of target nucleus with mass number $A$,
having mass $M_{A}$ and radius $R$;
${\bf p}_{t}$  and $E$ are the momentum and binding energy of the intranuclear target proton,
participating in the process (1). In our calculations we suppose that the proton and neutron densities
have the same radial shape $\rho(r)$. For it we have adopted an harmonic oscillator and a two-parameter
Fermi distributions for $^{12}$C and $^{184}$W target nuclei, correspondingly.

 As in Ref. [64], we will assume that the off-shell differential cross section\\
$d\sigma_{{\gamma}p\to {\Upsilon(1S)}p}(\sqrt{s^*},{\bf p}_{\Upsilon(1S)})/d{\bf p}_{\Upsilon(1S)}$
for $\Upsilon(1S)$ production in process (1) is the same as the corresponding on-shell cross section
$d\sigma_{{\gamma}p\to {\Upsilon(1S)}p}(\sqrt{s},{\bf p}_{\Upsilon(1S)})/d{\bf p}_{\Upsilon(1S)}$
determined for the off-shell kinematics of this process and in which the vacuum c.m. energy squared $s$,
given above by the formula (8), is replaced by the in-medium expression (29).
Then, the above-mentioned off-shell differential cross section is given by (cf. [64]):
%FORMULA (31)
\begin{equation}
\frac{d\sigma_{{\gamma}p \to {\Upsilon(1S)}p}(\sqrt{s^*},{\bf p}_{\Upsilon(1S)})}
{d{\bf p}_{\Upsilon(1S)}}=
\frac{\pi}{I_2(s^*,m_{\Upsilon(1S)},m_{p})E_{\Upsilon(1S)}}
\end{equation}
$$
\times
\frac{d\sigma_{{\gamma}p \to {\Upsilon(1S)}{p}}(\sqrt{s^*},\theta_{\Upsilon(1S)}^*)}{d{\bf \Omega}_{\Upsilon(1S)}^*}
\frac{1}{(\omega+E_t)}\delta\left[\omega+E_t-\sqrt{m_{p}^2+({\bf Q}+{\bf p}_t)^2}\right],
$$
where
%FORMULA (32)
\begin{equation}
I_2(s^*,m_{\Upsilon(1S)},m_{p})=\frac{\pi}{2}
\frac{\lambda(s^*,m_{\Upsilon(1S)}^{2},m_{p}^{2})}{s^*},
\end{equation}
%FORMULA (33)
\begin{equation}
\omega=E_{\gamma}-E_{\Upsilon(1S)}, \,\,\,\,{\bf Q}={\bf p}_{\gamma}-{\bf p}_{\Upsilon(1S)}.
\end{equation}
Here, the off-shell c.m. bottomonium angular distribution in process (1)
$d\sigma_{{\gamma}p \to {\Upsilon(1S)}{p}}(\sqrt{s^*},\theta_{\Upsilon(1S)}^*)/d{\bf \Omega}_{\Upsilon(1S)}^*$
as a function of the $\Upsilon(1S)$ production c.m. polar angle $\theta_{\Upsilon(1S)}^*$ is given, in
analogy with the $J/\psi$ meson photoproduction on the proton [64], by a standard expression:
%formula(34)
\begin{equation}
\frac{d\sigma_{{\gamma}p \to {\Upsilon(1S)}p}(\sqrt{s^*},\theta^*_{\Upsilon(1S)})}{d{\bf \Omega}_{\Upsilon(1S)}^*}=
a{\rm e}^{b_{\Upsilon(1S)}(t-t^+)}\sigma_{{\gamma}p \to {\Upsilon(1S)}p}(\sqrt{s^*}),
\end{equation}
where $\sigma_{{\gamma}p \to {\Upsilon(1S)}p}(\sqrt{s^*})$ is the off-shell total cross section
for bottomonium production in this process, $t$ is the momentum transfer to the proton squared
and $t^+$ is its maximum value, corresponding to the $t$ where the $\Upsilon(1S)$ is produced
at angle of 0$^{\circ}$ in the ${\gamma}p$ c.m. frame. According to [64], we can get that
%formula(35)
\begin{equation}
t-t^+=2p^*_{\gamma}p^*_{\Upsilon(1S)}(\cos{\theta^*_{\Upsilon(1S)}}-1).
\end{equation}
Here, the $\Upsilon(1S)$ c.m. momentum $p^*_{\Upsilon(1S)}$ is defined by Eq. (7), in which the free space invariant
collision energy squared $s$ should be replaced by the in-medium center-of-mass energy squared $s^*$ determined by
the formula (29) and
%formula(36)
\begin{equation}
p^*_{\gamma}=\frac{1}{2\sqrt{s^*}}\lambda(s^*,0,E_{t}^2-p_t^2).
\end{equation}
When the reaction ${\gamma}p \to {\Upsilon(1S)}p$ occurs on an on-shell target proton, then instead of the
in-medium collision energy squared $s^*$ and the quantity $E_{t}^2-p_t^2$, appearing in the equation (36),
we should use in it, respectively, the free space collision energy squared $s$ and the proton mass squared $m_p^2$.
The polar angle of bottomonium production in the ${\gamma}p$ c.m. system, $\theta^*_{\Upsilon(1S)}$, is expressed
through its polar production angle, $\theta_{\Upsilon(1S)}$, in the laboratory system by means of equation [64]:
%FORMULA (37)
\begin{equation}
\cos{\theta_{\Upsilon(1S)}^*}=\frac{p_{\gamma}p_{\Upsilon(1S)}\cos{\theta_{\Upsilon(1S)}}+
(E_{\gamma}^*E_{\Upsilon(1S)}^*-E_{\gamma}E_{\Upsilon(1S)})}{p_{\gamma}^*p_{\Upsilon(1S)}^*}.
\end{equation}
Here, the c.m. initial photon energy $E^*_{\gamma}$=$p^*_{\gamma}$.
The condition of normalization
%formula(38)
\begin{equation}
\int\limits_{4\pi}^{}a{\rm e}^{b_{\Upsilon(1S)}(t-t^+)}d{\bf \Omega}_{\Upsilon(1S)}^*=1
\end{equation}
gives for the parameter $a$ in Eq. (34) the following expression:
%formula(39)
\begin{equation}
a=\frac{p^*_{\gamma}p^*_{\Upsilon(1S)}b_{\Upsilon(1S)}}{\pi}\left[1-{\rm e}^{-4p^*_{\gamma}p^*_{\Upsilon(1S)}b_{\Upsilon(1S)}}\right]^{-1}.
\end{equation}
%%%%%%%%%%%%%%%%%%%%%%%%%%%%%%%%%%%%%%%%%%%%%%%%%%%%%%%%%%%
\begin{figure}[htb]
\begin{center}
\includegraphics[width=16.0cm]{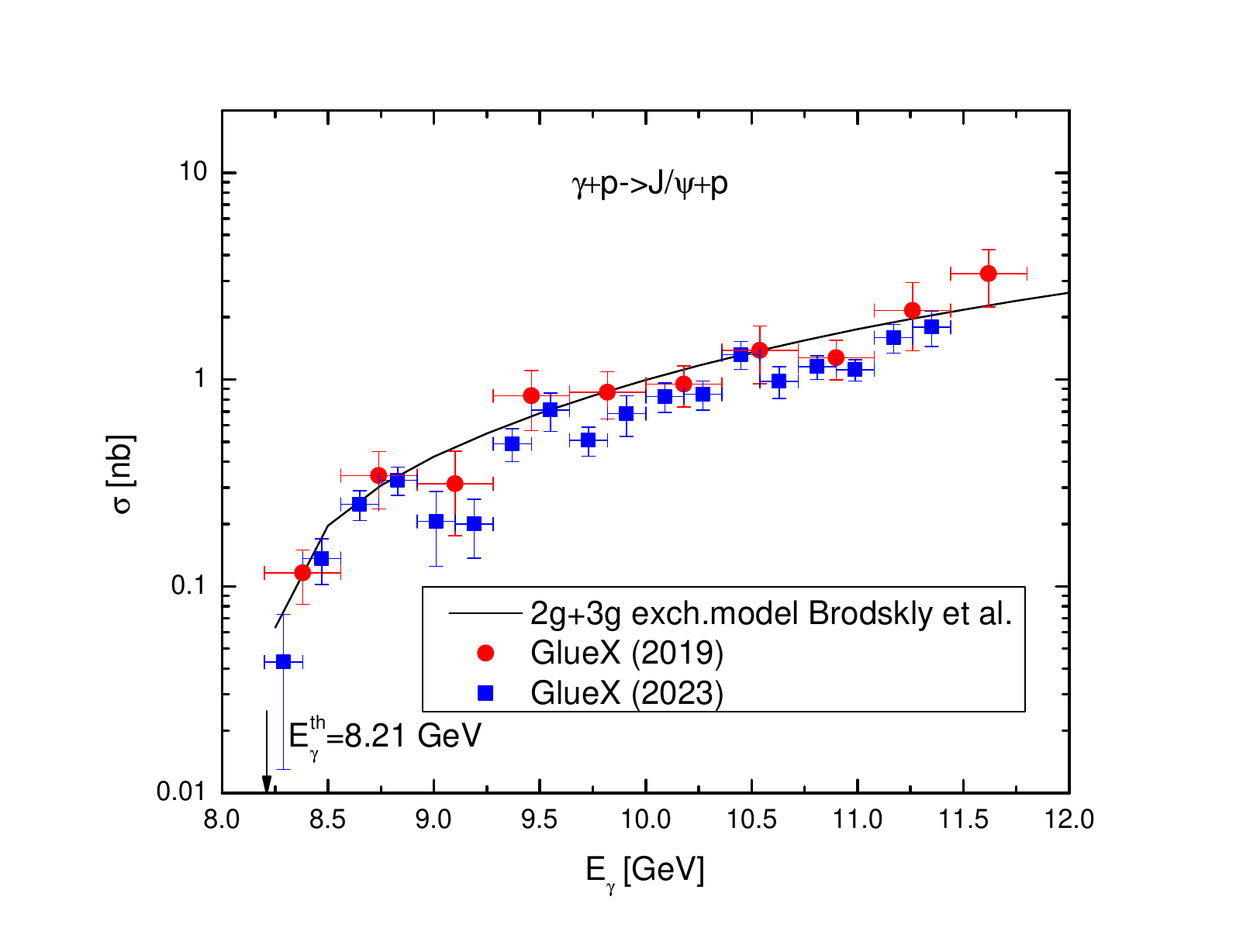}
\vspace*{-2mm} \caption{(Color online.) The total cross section for the reaction
${\gamma}p \to {J/\psi}p$ as a function of the photon energy $E_{\gamma}$.
Solid curve is the incoherent sum of the two calculations, performed
on the basis, respectively, of the two gluon and three gluon exchange model [82].
The GlueX (2019) and GlueX (2023) experimental data are from Refs. [81] and [83], respectively.
The arrow indicates the threshold energy of 8.21 GeV
for direct non-resonant charmonium photoproduction on a free target proton at rest.}
\label{void}
\end{center}
\end{figure}
%%%%%%%%%%%%%%%%%%%%%%%%%%%%%%%%%%%%%%%%%%%%%%%%%%%%%%%%%%%%%%

We focus now on the "in-medium" total cross section $\sigma_{{\gamma}p \to {\Upsilon(1S)}p}(\sqrt{s^*})$ for
$\Upsilon(1S)$ production in reaction (1). According to the aforementioned, it is equivalent to the vacuum cross
section $\sigma_{{\gamma}p \to {\Upsilon(1S)}p}(\sqrt{s})$, in which the vacuum c.m. energy squared $s$, presented
by the formula (8), is replaced by the in-medium one $s^*$, given by the expression (29).
For the free total cross section $\sigma_{{\gamma}p \to {\Upsilon(1S)}p}(\sqrt{s})$ no data are available presently
in the considered threshold energy region $W \le 11.4$ GeV
\footnote{$^)$It should be noticed that the experimental data on $\Upsilon(1S)$ meson production in the reaction
${\gamma}p \to \Upsilon(1S)p$ are available only at high photon-proton center-of-mass energies
$W=\sqrt{s} > 60$ GeV  (cf. [63]).}$^)$,
where the masses of the predicted $P_b$ states are concentrated and where they can be observed in the ${\gamma}p$ reactions [63]. So, we have to rely on some theoretical predictions and estimates for it,
existing in the literature in this energy region. Thus, in Ref. [63] for the latter cross section at near-threshold
photon energies we have used the following expression, based on an analysis of the data on the production of
$\Upsilon(1S)$ and $J/\psi$ mesons in ${\gamma}p$ collisions in the kinematic range $80 < W < 160$ GeV,
performed by the ZEUS Collaboration at HERA [79]:
%formula(40)
\begin{equation}
 \sigma_{{\gamma}p \to {\Upsilon(1S)}p}(\sqrt{s})=5\cdot10^{-3}\cdot
 \sigma_{{\gamma}p \to {J/\psi}p}(\sqrt{{\tilde s}}),
\end{equation}
where the quantity $\sigma_{{\gamma}p \to {J/\psi}p}(\sqrt{{\tilde s}})$ represents the free total cross
section of the process ${\gamma}p \to {J/\psi}p$ and
the center-of-mass collision energies $\sqrt{s}$ and $\sqrt{{\tilde s}}$ are linked by the relation:
%formula(41)
\begin{equation}
\sqrt{{\tilde s}}=\sqrt{s}-m_{\Upsilon(1S)}+m_{J/\psi}.
\end{equation}
Here, the quantity $m_{J/\psi}$ is the vacuum $J/\psi$ meson mass.
At low initial photon energies $\sqrt{s} \le 11.4$ GeV of our interest,
the c.m. energy $\sqrt{{\tilde s}} \le 5.04$ GeV. The latter
corresponds, as is easy to see, to the laboratory photon energy domain $\le$ 13.05 GeV.
For the free total cross section $\sigma_{{\gamma}p \to {J/\psi}p}({\sqrt{{\tilde s}}})$
in this domain we have adopted the following parametrization [63, 80] of the available here
experimental data [81] on it from the GlueX experiment from 2019,
based on the predictions of the two gluon and three gluon exchange model [82] near threshold:
%formula(42)
\begin{equation}
\sigma_{{\gamma}p \to {J/\psi}p}({\sqrt{{\tilde s}}})= \sigma_{2g}({\sqrt{{\tilde s}}})+
\sigma_{3g}({\sqrt{{\tilde s}}}),
\end{equation}
where 2$g$ and 3$g$ exchanges cross sections $\sigma_{2g}({\sqrt{{\tilde s}}})$ and
$\sigma_{3g}({\sqrt{{\tilde s}}})$ are given in Ref. [63] by formulas (14) and (15), respectively.
Fig 2 shows that the new GlueX near-threshold data [83] from 2023 are also reasonably well fitted by
the combination (42), except of a sharp dip structure observed in the total $\gamma+p \to J/\psi+p$ cross section
in the vicinity of $E_{\gamma}=9.1$ GeV. It is interesting to note that the possibility that this dip structure
is produced due to the destructive interference involving an $S$-wave $P_c(4312)^+$ resonance and associated
non-resonance background was considered in recent publication [84]. The authors of Ref. [85] considered also a wide array of physics possibilities that are compatible with present $J/\psi$ photoproduction data (in particular, a contribution
from open charm intermediate ${\Lambda_c}{\bar D}$ and ${\Lambda_c}{\bar D}^*$ states
\footnote{$^)$See Ref. [83] as well.}$^)$)
and which need to be disentangled.
%%%%%%%%%%%%%%%%%%%%%%%%%%%%%%%%%%%%%%%%%%%%%%%%%%%%%%%%%%%
\begin{figure}[htb]
\begin{center}
\includegraphics[width=16.0cm]{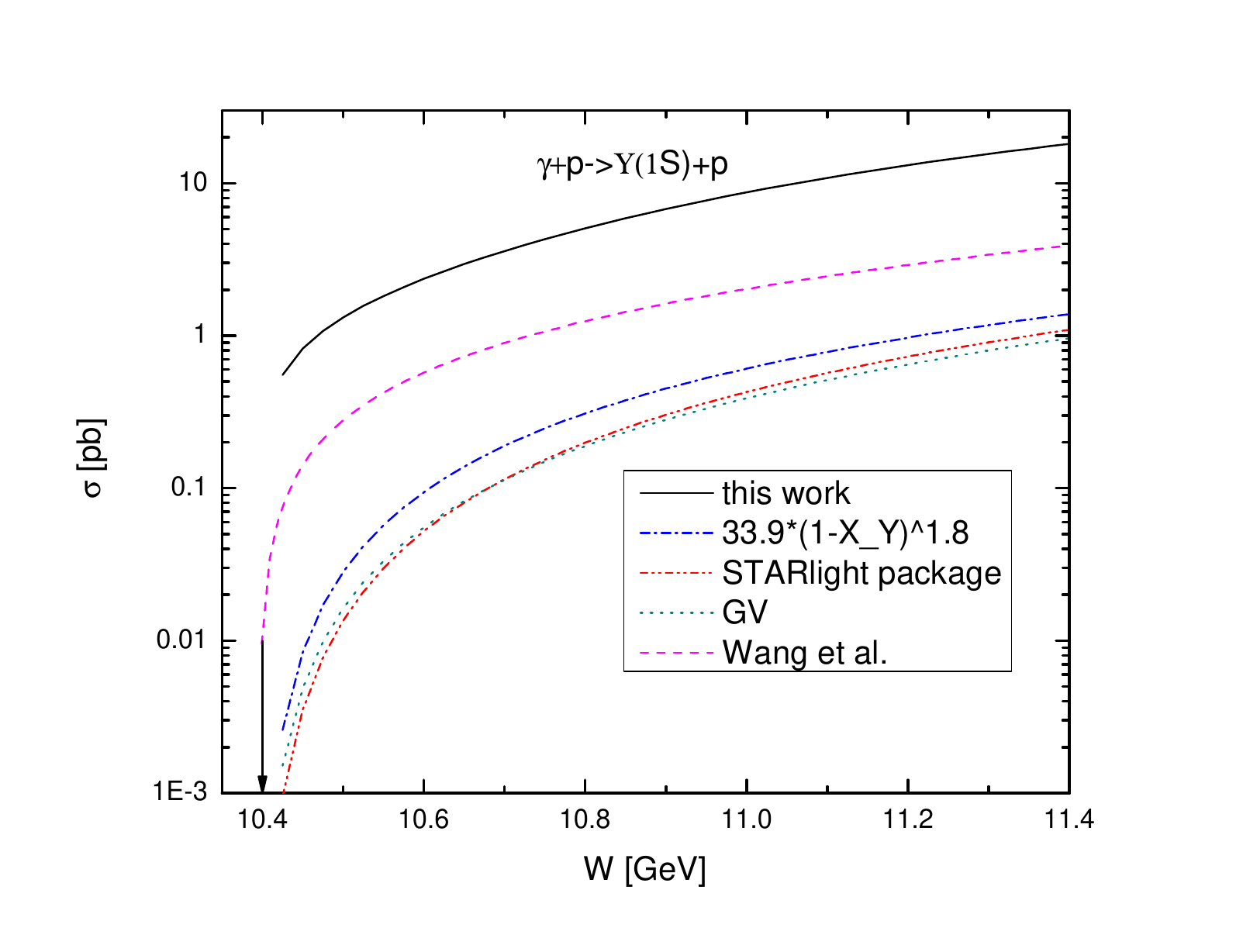}
\vspace*{-2mm} \caption{(Color online.) The non-resonant total cross section for the background reaction
${\gamma}p \to {\Upsilon(1S)}p$ as a function of the center-of-mass energy $W=\sqrt{s}$
of the photon--proton collisions.
Solid, dashed, dashed-dotted, dashed-dotted-dotted and dotted curves are calculations by (40)--(42),
from Ref. [61] within the dipole Pomeron model [86], using Eqs. (43), (45) and (46), respectively.
The arrow indicates the center-of-mass threshold energy
for direct $\Upsilon(1S)$ photoproduction on a free target proton being at rest.}
\label{void}
\end{center}
\end{figure}
%%%%%%%%%%%%%%%%%%%%%%%%%%%%%%%%%%%%%%%%%%%%%%%%%%%%%%%%%%%
  The results of calculations by Eqs. (40)--(42) of the non-resonant total cross section of the reaction
${\gamma}p \to {\Upsilon(1S)}p$ at relative "low" c.m. energies $W \le 11.4$ GeV, corresponding to the EicC energy
coverage [65, 66], are shown in Fig. 3 (solid curve).
In this figure we also show the predictions from Ref. [61] obtained within the dipole Pomeron model [86] (dashed curve),
from the recently proposed parametrization [87]
%formula(43)
\begin{equation}
 \sigma_{{\gamma}p \to {\Upsilon(1S)}p}(\sqrt{s})=33.9(1-x_{\Upsilon})^{1.8}~[\rm pb],
\end{equation}
where $x_{\Upsilon}$ is defined as
%formula(44)
\begin{equation}
  x_{\Upsilon}=(s_{\rm th}-m^2_p)/(s-m^2_p), \,\,\,s_{\rm th}=(m_p+m_{\Upsilon(1S)})^2
\end{equation}
(dashed-dotted curve). In addition, the predictions from the current parametrization
%formula(45)
\begin{equation}
 \sigma_{{\gamma}p \to {\Upsilon(1S)}p}(\sqrt{s})=6.4\left(1-\frac{s_{\rm th}}{s}\right)^2\cdot(\sqrt{s})^{0.74}~[\rm pb],
\end{equation}
employed in the STARlight Monte Carlo simulation program package [88] to simulate the $\Upsilon(1S)$ production in ultra-peripheral collisions of relativistic ions, and from the parametrization
%formula(46)
\begin{equation}
 \sigma_{{\gamma}p \to {\Upsilon(1S)}p}(\sqrt{s})=\left(\frac{ef_{\Upsilon(1S)}}{m_{\Upsilon(1S)}}\right)^2
 \frac{N}{2\sqrt{s}p^*_{\gamma}}\left(\frac{p^*_{\Upsilon(1S)}}{p^*_{\gamma}}\right)
 \left(1-\frac{\nu_{\rm el}}{\nu}\right)^{b_{\rm el}}\left(\frac{\nu}{\nu_{\rm el}}\right)^{a_{\rm el}},
\end{equation}
where
%formula(47)
\begin{equation}
\nu=(s-m^2_p-m^2_{\Upsilon(1S)})/2,\,\,\,\nu_{\rm el}=m_pm_{\Upsilon(1S)},\,\,\,a_{\rm el}=1.38,\,\,\,
b_{\rm el}=1.27,\,\,\,N=0.014,\,\,\,f_{\Upsilon(1S)}=238~{\rm MeV},
\end{equation}
proposed by Gryniuk and Vanderhaeghen (GV) [89]
\footnote{$^)$It should be pointed out that for parameters $a_{\rm el}$, $b_{\rm el}$ and $N$, entering into
Eq. (46), we have used in our calculations their central values, presented in Eq. (47). The central value of the
parameter N is taken from Ref. [62].}$^)$,
are also shown in Fig. 3 (dashed-dotted-dotted and dotted curves, respectively).
This figure shows that the deviations between our parametrization (40)-(42) and those given above by
Eqs. (43)-(47) are large at all considered energies, which are covered by the proposed EicC.
In particular, at photon energies around 11 GeV, at which the pentaquarks $P_b$ should appear, our parametrization (40)--(42) is substantially larger (by factors of about 5 and 15, respectively) than the results obtained
in Ref. [61] using the dipole Pomeron model [86] and the parametrizations (43), (45), (46).
The parametrization (45), adopted in the STARlight package, overlaps with the GV parametrization (46) and they
are slightly smaller than the current parametrization (43).
Therefore, in our subsequent calculations we will use the empirical formulas (40)--(42) and (45) as guidelines for
a reasonable estimation of the upper and lower limits of the non-resonant contribution under the expected pentaquark peaks.
%%%%%%%%%%%%%%%%%%%%%%%%%%%%%%%%%%%%%%%%%%%%%%%%%%%%%%%%%%%
\begin{figure}[htb]
\begin{center}
\includegraphics[width=16.0cm]{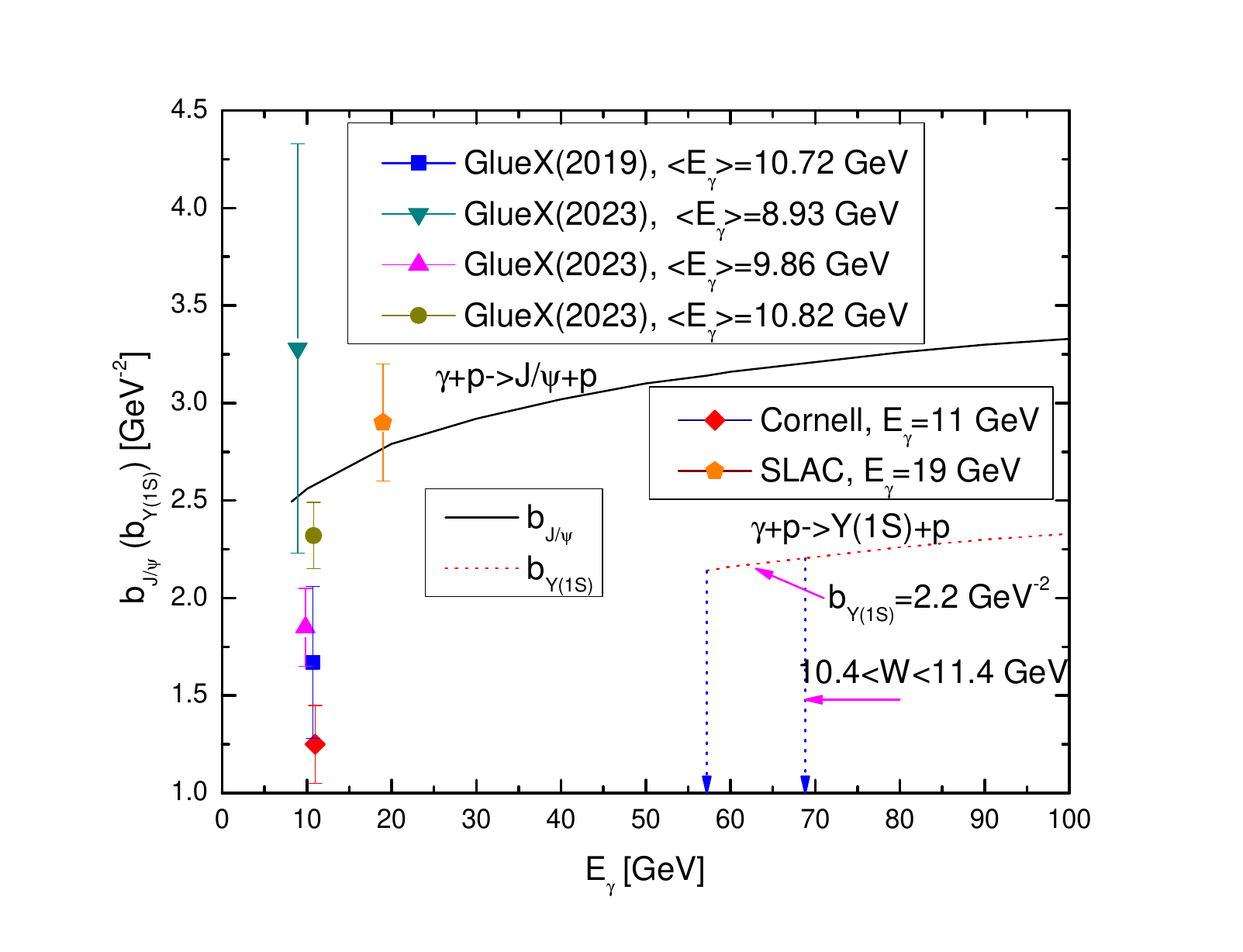}
\vspace*{-2mm} \caption{(Color online.) The photon energy dependencies of the exponential slopes $b_{J/\psi}$ (solid curve) and $b_{\Upsilon(1S)}$ (dotted curve)
for $J/\psi$ meson and 1S-bottomonium photoproduction in the free space processes ${\gamma}p \to {J/\psi}p$
and ${\gamma}p \to {\Upsilon(1S)}p$ given by Eqs. (49) and (48) (49), respectively.
The experimental data points for the slope $b_{J/\psi}$ in the low energy region near the ${J/\psi}p$ threshold
are from Refs. [81, 83, 91, 92] (see, also, the text). The vertical dotted arrows indicate the
range of initial photon energies in the laboratory frame near the $\Upsilon(1S)p$ threshold, considered in the present work.}
\label{void}
\end{center}
\end{figure}
%%%%%%%%%%%%%%%%%%%%%%%%%%%%%%%%%%%%%%%%%%%%%%%%%%%%%%%%%%%

In order to calculate the differential cross section (34) one should know an exponential $t$-slope parameter
$b_{\Upsilon(1S)}$ for the non-resonant process ${\gamma}p \to {\Upsilon(1S)}p$ in the near-threshold energy
region. In free space, this slope was fitted to high-energy data in [90] and found to have a smaller value
than for $J/\psi$ (cf. Ref. [76]):
%formula(48)
\begin{equation}
b_{\Upsilon(1S)}(s) \approx b_{J/\psi}(s)-1~{\rm GeV}^{-2}.
\end{equation}
For the c.m. energy behavior of the diffraction slope $b_{J/\psi}(s)$ we use the standard Regge form [76, 90]:
%formula(49)
\begin{equation}
b_{J/\psi}(s)=b_0+2{\alpha}^{\prime}(0){\rm ln}\left(\frac{s}{s_0}\right),
\end{equation}
where the parameters ${\alpha}^{\prime}(0)=0.171~{\rm GeV}^{-2}$ and $b_0=1.54~{\rm GeV}^{-2}$ were fitted in [90]
to data on $J/\psi$ photoproduction with $s_0=1$~GeV$^2$.
Fig. 4 shows the results of calculations by (49) and (48), (49) for the slope parameters $b_{J/\psi}$ (solid
curve) and $b_{\Upsilon(1S)}$ (dotted curve) at incident photon beam energies below 100 GeV. It is seen that the $t$-slope $b_{\Upsilon(1S)}$ has a weak energy dependence and
$b_{\Upsilon(1S)}\approx$ 2.2 GeV$^{-2}$ in the photon energy range of $57.2~{\rm GeV} \le E_{\gamma} \le 68.8~{\rm GeV}$ (or in the c.m. energy range of $10.4~{\rm GeV} \le W \le 11.4~{\rm GeV}$) of interest.
We will use this value in our calculations. It is worth noting that the extrapolation of the simple fit (49) of
the high-energy data to the ${J/\psi}p$ threshold energies (to the energies around 10 GeV) is also compatible with
available here data for an exponential $J/\psi$ $t$-slope $b_{J/\psi}$, namely: with the Cornel result at
$E_{\gamma}=11$ GeV of 1.25$\pm$0.20 GeV$^{-2}$ [91], the SLAC result at $E_{\gamma}=19$ GeV
of 2.9$\pm$0.3 GeV$^{-2}$ [92], the GlueX(2019) result of 1.67$\pm$0.39 GeV$^{-2}$
at average energy of 10.72 GeV [81] and the GlueX(2023) results of 3.28$\pm$1.05 GeV$^{-2}$, 1.85$\pm$0.20 GeV$^{-2}$, 2.32$\pm$0.17 GeV$^{-2}$ at average photon energies of 8.93 GeV, 9.86 GeV, 10.82 GeV, respectively.
The GlueX(2023) results for the $t$-slope $b_{J/\psi}$ were calculated using the relation
$b_{J/\psi}\approx4/m^2_s$, which can be easily obtained by transforming the dipole $t$-dependence for the
differential cross section of the reaction ${\gamma}p \to {J/\psi}p$ of the form
$[d\sigma/dt(0)]/(1-t/m^2_s)^4$ with a mass scale $m_s$ of 1.105$\pm$0.168 GeV, 1.472$\pm$0.075 GeV, 1.313$\pm$0.049 GeV
measured [83] at these energies to an exponential form (34).
Therefore, one may hope that the uncertainty of $b_{\Upsilon(1S)}$ is not large, and it does not affect strongly
the uncertainty of the differential cross section (34).
It should be pointed out that the differential cross section of the reaction ${\gamma}p \to {J/\psi}p$
was also very recently measured in the $J/\psi$-007 experiment [93]
as a function of the photon energy in the range of $9.1~{\rm GeV} \le E_{\gamma} \le 10.6~{\rm GeV}$.
But the $t$-slope $b_{J/\psi}$ was not determined in [93].
The collected here differential $J/\psi$ photoproduction data together with those from the GlueX experiments [81, 83]
were used in the determination of such fundamental quantities, characterizing the proton internal structure, as the gluonic gravitational form factors of the proton [93--97], its gluonic mass radius [95--99]
\footnote{$^)$It should be noted that a way to access the proton internal structure via the near-threshold
heavy quarkonium photoproduction has been proposed by Kharzeev and collaborators in Refs. [100, 101].}$^)$
.
The process to determine gluonic gravitational form factors of the proton from near-threshold vector meson photoproduction is currently under active discussion (cf. Refs. [93--99, 102]).
It is emphasized that more measurements with higher quality, especially at large $|t|$, at different photon energies
are of crucial importance to further improve their (and the gluonic mass radius of the proton) extraction from the near-threshold $J/\psi$ and $\Upsilon(1S)$ production measurements. We expect that such measurements will be performed in the future at JLab 12 and 24 GeV energies [103--105] and at the future electron-ion colliders [66, 68, 70, 106, 107].
At the same time, the near-threshold $\Upsilon(1S)$ photoproduction experiments are more beneficial than the $J/\psi$
ones for their measurement since in these experiments the larger kinematic coverage of momentum transfer (due to large mass difference between $J/\psi$ and $\Upsilon(1S)$) will be realized, which gives a strong argument to apply perturbative QCD to understand the
near-threshold heavy quarkonium production [108]. Moreover, they will also allow the search for the predicted but not yet observed hidden-bottom partners of the  $P^+_c$ signals observed in the ${J/\psi}p$ mass spectrum by the LHCb Collaboration.

At the incident photon energies of interest the $\Upsilon(1S)$ mesons are produced at very small
laboratory polar angles (see Eq. (10)). Therefore, we will calculate their momentum
distributions from considered target nuclei
for the laboratory solid angle
${\Delta}{\bf \Omega}_{\Upsilon(1S)}$ = $0^{\circ} \le \theta_{\Upsilon(1S)} \le 10^{\circ}$,
and $0 \le \varphi_{\Upsilon(1S)} \le 2{\pi}$.
Then, integrating the full differential cross section (26) over this solid angle,
we can represent the momentum-dependent differential cross section for bottomonium
production from the direct non-resonant processes (1) and (2) into this solid angle
in the form (cf. [64]):
%formula(50)
\begin{equation}
\frac{d\sigma_{{\gamma}A\to {\Upsilon(1S)}X}^{({\rm dir})}
(p_{\gamma},p_{\Upsilon(1S)})}{dp_{\Upsilon(1S)}}=
2{\pi}I_{V}[A,\sigma_{{\Upsilon(1S)}N}]
\int\limits_{\cos10^{\circ}}^{1}d\cos{{\theta_{\Upsilon(1S)}}}
\left<\frac{d\sigma_{{\gamma}p\to {\Upsilon(1S)}{p}}(p_{\gamma},
p_{\Upsilon(1S)},\theta_{\Upsilon(1S)})}{dp_{\Upsilon(1S)}d{\bf \Omega}_{\Upsilon(1S)}}\right>_A.
\end{equation}
For our purposes, we also need the $\Upsilon(1S)$ energy spectrum
$d\sigma_{{\gamma}p \to {\Upsilon(1S)}p}[\sqrt{s},p_{\Upsilon(1S)}]/ dE_{\Upsilon(1S)}$ from
the ${\gamma}p \to {\Upsilon(1S)}p$ reaction, taking place on a free target proton at rest, as a function of
the $\Upsilon(1S)$ total energy $E_{\Upsilon(1S)}$ belonging to the kinematically allowed interval (18).
Accounting for Eq. (31) and using the results given in Ref. [64], it was calculated to be:
%formula(51)
\begin{equation}
\frac{d\sigma_{{\gamma}p \to {\Upsilon(1S)}p}[\sqrt{s},p_{\Upsilon(1S)}]}{dE_{\Upsilon(1S)}}=
\left(\frac{2{\pi}\sqrt{s}}{p_{\gamma}p^{*}_{\Upsilon(1S)}}\right)
\frac{d\sigma_{{\gamma}p \to {\Upsilon(1S)}p}[\sqrt{s},\theta_{\Upsilon(1S)}^*(x_0)]}{d{\bf \Omega}_{\Upsilon(1S)}^*}~{\rm for}
~E^{(2)}_{\Upsilon(1S)}(0^{\circ}) \le E_{\Upsilon(1S)} \le E^{(1)}_{\Upsilon(1S)}(0^{\circ}),
\end{equation}
where
%formula(52)
\begin{equation}
x_0=\frac{[p^2_{\gamma}+p^2_{\Upsilon(1S)}+m^2_{p}-(\omega+m_p)^2]}{2p_{\gamma}p_{\Upsilon(1S)}},\,\,\,\,\,
p_{\Upsilon(1S)}=\sqrt{E^2_{\Upsilon(1S)}-m^2_{\Upsilon(1S)}}
\end{equation}
and the quantity $\cos{\theta_{\Upsilon(1S)}^*(x_0)}$ is defined by Eq. (37), in which one has to perform the
replacement: $\cos{\theta_{\Upsilon(1S)}} \to x_0$, and the $\Upsilon(1S)$ and the photon c.m. momenta
$p^*_{\Upsilon(1S)}$ and $p^*_{\gamma}$ are defined by Eqs. (7) and (36), respectively. In the latter
one needs also to make the substitutions: $E_t \to m_p$, $p_t \to 0$ and $s^* \to s$.
The expressions (51), (52) will be used by us for calculating the free space $\Upsilon(1S)$ energy spectrum
for incident resonant photon beam energies of 64.952, 65.484 and 65.544 GeV (see below).

\subsection*{2.2. Two-step resonant $\Upsilon(1S)$ production processes}

At photon beam c.m. energies below 11.4 GeV, incident photons can produce the predicted [42], but non-observed
yet non-strange charged $P^+_b(11080)$, $P^+_b(11125)$, $P^+_b(11130)$
and neutral $P^0_b(11080)$, $P^0_b(11125)$, $P^0_b(11130)$ hidden-bottom pentaquark resonances with quark
structures $|P^+_b>=|uudb{\bar b}>$ and $|P^0_b>=|uddb{\bar b}>$ and
with pole masses $M_{b1}=11080$ MeV, $M_{b2}=11125$ MeV, $M_{b3}=11130$ MeV, respectively,
directly in the first inelastic collisions with intranuclear protons and neutrons:
%formula(53)
\begin{eqnarray}
{\gamma}+p \to P^+_b(11080),\nonumber\\
{\gamma}+p \to P^+_b(11125),\nonumber\\
{\gamma}+p \to P^+_b(11130);
\end{eqnarray}
%formula(54)
\begin{eqnarray}
{\gamma}+n \to P^0_b(11080),\nonumber\\
{\gamma}+n \to P^0_b(11125),\nonumber\\
{\gamma}+n \to P^0_b(11130).
\end{eqnarray}
Then the produced pentaquark resonances can decay into the final states ${\Upsilon(1S)}p$ and ${\Upsilon(1S)}n$,
which will additionally contribute to the $\Upsilon(1S)$ yield in the ($\gamma$,$\Upsilon(1S)$) reactions
on protons and nuclei:
%formula(55)
\begin{eqnarray}
P^+_b(11080) \to \Upsilon(1S)+p,\nonumber\\
P^+_b(11125) \to \Upsilon(1S)+p,\nonumber\\
P^+_b(11130) \to \Upsilon(1S)+p;
\end{eqnarray}
%formula(56)
\begin{eqnarray}
P^0_b(11080) \to \Upsilon(1S)+n,\nonumber\\
P^0_b(11125) \to \Upsilon(1S)+n,\nonumber\\
P^0_b(11130) \to \Upsilon(1S)+n.
\end{eqnarray}
As before in Ref. [63], we assume in this exploratory study that the $P^+_{bi}$ and $P^0_{bi}$ states
\footnote{$^)$Here, $i=$1, 2, 3. $P^+_{b1}$, $P^+_{b2}$, $P^+_{b3}$ and $P^0_{b1}$, $P^0_{b2}$, $P^0_{b3}$
stand for $P^+_b(11080)$, $P^+_b(11125)$, $P^+_b(11130)$ and $P^0_b(11080)$, $P^0_b(11125)$, $P^0_b(11130)$, respectively.}$^)$
with the possible spin-parity quantum numbers $J^P=(1/2)^-$ for $P^+_{b1}$ and $P^0_{b1}$,
$J^P=(1/2)^-$ for $P^+_{b2}$ and $P^0_{b2}$, $J^P=(3/2)^-$ for $P^+_{b3}$ and $P^0_{b3}$
\footnote{$^)$Which might be assigned to them within
the hadronic molecular scenario for their internal structure [42].}$^)$
have the same total widths $\Gamma_{bi}$ and they are the same as for their hidden-charm partners $P^+_c(4312)$, $P^+_c(4440)$ and $P^+_c(4457)$, i.e., $\Gamma_{b1}=9.8$ MeV, $\Gamma_{b2}=20.6$ MeV, $\Gamma_{b3}=6.4$ MeV [9].
In addition, we will suppose that the branching ratios
$Br[P^+_{bi} \to {\Upsilon(1S)}p]$ and $Br[P^0_{bi} \to {\Upsilon(1S)}n]$ ($i=$1, 2, 3) of the decays (55) and (56)
are equal to each other [42] and will adopt, relying on the similarity of the behaviors and decay properties of the
$P_b$ and $P_c$ systems, for each individual $P_b$ state the same three conservative options:
0.25, 0.5 and 1\% as those used in Ref. [64] for the $P^+_{c} \to {J/\psi}p$ decays (cf. Ref. [66]), and additional one
with reduced value of 0.125\% of these ratios in order to see better their impact
on the resonant $\Upsilon(1S)$ yield in ${\gamma}$$p$ $\to {\Upsilon(1S)}p$,
${\gamma}$$^{12}$C $\to {\Upsilon(1S)}X$ and ${\gamma}$$^{184}$W $\to {\Upsilon(1S)}X$ reactions.
This choice for the branching fractions $Br[P^+_{c} \to {J/\psi}p]$ of the $P^+_{c} \to {J/\psi}p$ decays
is caused by the fact that the current upper and lower limits on them sit at the levels of 1\% [64, 85, 109] and
0.05\% $\sim$ 0.5\% [110], respectively. Moreover, it is also supported by the recent investigations, performed in particular in Refs. [45, 111]. Thus, using the holographic QCD in the heavy quark limit for the description of
the charm pentaquark states, the authors of Ref. [45] calculated the branching ratios
$Br[P^+_{c}(4312) \to {J/\psi}p]$, $Br[P^+_{c}(4440) \to {J/\psi}p]$ and $Br[P^+_{c}(4457) \to {J/\psi}p]$
to be 0.3\%, 1.8\% and 0.9\%, correspondingly. Combining the effective Lagrangian and the Bethe-Salpeter framework,
the partial decay width of $P^+_{c}(4312)$ resonance to ${J/\psi}p$ was calculated to be 0.17 MeV in Ref. [111].
It is $\approx$ 1.7\% of the total width $\approx$ 9.8 MeV reported by the LHCb collaboration [9].
These values are close to the indicated above upper bound of about 1\% for the branching fractions $Br[P^+_{c} \to {J/\psi}p]$. On the other hand, the obtained in Ref. [46] results for the partial and total decay widths of the bottom
pentaquarks $P_b$ indicate that the branching fractions of their decays to the ${\Upsilon(1S)}p$ mode are relatively
small and they are of about 0.01--0.06\%, which is close to the lower bound of $\sim$ 0.05\% empirically argued in [110]
for the branching ratios of the $P^+_{c} \to {J/\psi}p$ decays. In addition to the aforesaid theoretical activities,
it is worth mentioning that the range
of 0.5\%--5\% was chosen in Ref. [62] for the branching fractions of the $P_b \to {\Upsilon(1S)}p$ decays.
In line with the aforementioned, one may conclude that the chosen in the present paper options for these fractions
seem to be more realistic than those of 2, 3, 5 and 10\% employed for them in Ref. [63].

According to Ref. [63], we assume that the $P^+_{bi}$ and $P^0_{bi}$ in-medium spectral functions
$S_{bi}^+(\sqrt{s^*},\Gamma_{bi})$ and $S_{bi}^0(\sqrt{s^*},\Gamma_{bi})$
are described by the non-relativistic Breit-Wigner distributions
\footnote{$^)$To simplify our calculations we ignore the modification of the
$P_{bi}^{+}$ and $P_{bi}^{0}$ masses and total decay widths in the nuclear matter in the present study.}$^)$
:
%formula(57)
\begin{equation}
S_{bi}^+(\sqrt{s^*},\Gamma_{bi})=S_{bi}^0(\sqrt{s^*},\Gamma_{bi})=
\frac{1}{2\pi}\frac{\Gamma_{bi}}{(\sqrt{s^*}-M_{bi})^2+({\Gamma}_{bi})^{2}/4},\,\,\,i=1,2,3;
\end{equation}
where $\sqrt{s^*}$ is the total ${\gamma}N$ c.m.s. energy defined by Eq. (29).
At first, we consider the bottomonium production in the production/decay sequences (53)/(55), taking place on a bound in the target nucleus proton. According to [63, 64], the in-medium total cross sections
$\sigma_{{\gamma}p \to P_{bi}^+ \to {\Upsilon(1S)}p}(\sqrt{s^*},\Gamma_{bi})$ ($i=$1, 2, 3)
for resonant $\Upsilon(1S)$ production in these sequences can be represented as follows:
%FORMULA (58)
\begin{equation}
\sigma_{{\gamma}p \to P_{bi}^+ \to {\Upsilon(1S)}p}(\sqrt{s^*},\Gamma_{bi})=
\sigma_{{\gamma}p \to P_{bi}^+}(\sqrt{s^*},\Gamma_{bi})\theta[\sqrt{s^*}-(m_{\Upsilon(1S)}+m_{p})]
Br[P_{bi}^+ \to {\Upsilon(1S)}p].
\end{equation}
Here, $\theta(x)$ is the step function and
the quantities $\sigma_{{\gamma}p \to P_{bi}^+}(\sqrt{s^*},\Gamma_{bi})$ are the total cross sections for
production of the $P_{bi}^+$ resonances in reactions (53).
These cross sections can be described, using the spectral functions $S_{bi}^+(\sqrt{s^*},\Gamma_{bi})$
of resonances and knowing the branching fractions $Br[P^+_{bi} \to {\gamma}p]$ of their decays to
the ${\gamma}p$ mode, as follows [63, 64]:
%formula(59)
\begin{equation}
\sigma_{{\gamma}p \to P_{bi}^+}(\sqrt{s^*},\Gamma_{bi})=
f_{bi}\left(\frac{\pi}{p^*_{\gamma}}\right)^2
Br[P_{bi}^+ \to {\gamma}p]S_{bi}^+(\sqrt{s^*},\Gamma_{bi})\Gamma_{bi}, \,\,i=1,2,3,
\end{equation}
where the in-medium c.m. 3-momentum in the incoming ${\gamma}p$ channel, $p^*_{\gamma}$,
is defined above by the formula (36) and the ratios of the spin factors $f_{bi}$ are $f_{b1}=1$, $f_{b2}=1$, $f_{b3}=2$.
Inspection of Eqs. (58), (59) shows that to estimate the $\Upsilon(1S)$ production cross sections
from the production/decay chains (53)/(55), proceeding on a vacuum proton
(or on a bound in a nuclear medium proton), one needs to know yet the branching ratios
$Br[P^+_{bi} \to {\gamma}p]$ ($i=$1, 2, 3) of the $P^+_{bi}$ decays to the ${\gamma}p$ channel.
Assuming that the decays of the $P^+_{bi}$ states to ${\Upsilon(1S)}p$ are dominated by the lowest partial
waves with relative orbital angular momentum $L=0$ and using the vector-meson dominance model, we expressed in Ref. [63]
(see formula (39)) this branching ratios via the branching fractions $Br[P_{bi}^+ \to {\Upsilon(1S)}p]$.

In view of the aforesaid and Eq. (58), the free total cross sections
$\sigma_{{\gamma}p \to P_{bi}^+ \to {\Upsilon(1S)}p}(\sqrt{s},\Gamma_{bi})$ ($i=$1, 2, 3)
for bottomonium production in the two-step processes (53)/(55), taking place on the target proton being at rest,
can be represented in the form:
%FORMULA (60)
\begin{equation}
\sigma_{{\gamma}p \to P_{bi}^+ \to {\Upsilon(1S)}p}(\sqrt{s},\Gamma_{bi})=
\sigma_{{\gamma}p \to P_{bi}^+}(\sqrt{s},\Gamma_{bi})\theta[\sqrt{s}-(m_{\Upsilon(1S)}+m_{p})]
Br[P_{bi}^+ \to {\Upsilon(1S)}p],
\end{equation}
in which the vacuum cross sections $\sigma_{{\gamma}p \to P_{bi}^+}(\sqrt{s},\Gamma_{bi})$ are described
by Eq. (59), in which one should make the replacement $s^* \to s$ and determine the c.m. incident photon
momentum $p^*_{\gamma}$ as follows:
%FORMULA(61)
\begin{equation}
p_{\gamma}^*=\frac{1}{2\sqrt{{s}}}\lambda({s},0,m_{p}^2).
\end{equation}

Evidently, the in-medium total cross sections
$\sigma_{{\gamma}n \to P_{bi}^0 \to {\Upsilon(1S)}n}(\sqrt{s^*},\Gamma_{bi})$ ($i=$1, 2, 3)
for resonant $\Upsilon(1S)$ production in the production/decay sequences (54)/(56) can be described by the
expressions analogous to those of Eqs. (58), (59) obtained for the case of proton target. Considering
the relation $Br[P^0_{bi} \to {\Upsilon(1S)}n]=Br[P^+_{bi} \to {\Upsilon(1S)}p]$ [42] and supposing
that the decays of the $P^0_{bi}$ states to ${\Upsilon(1S)}n$ are also dominated, as in the case of
$P^+_{bi}$ resonances, by the lowest partial waves with relative orbital angular momentum $L=0$,
we can get within the vector-meson dominance model that [63]
%FORMULA (62)
\begin{equation}
Br[P_{bi}^0 \to {\gamma}n]=Br[P_{bi}^+ \to {\gamma}p],\,\,\, i=1, 2, 3.
\end{equation}
With this and Eq. (57), we obtain
%FORMULA (63)
\begin{equation}
\sigma_{{\gamma}n \to P_{bi}^0 \to {\Upsilon(1S)}n}(\sqrt{s^*},\Gamma_{bi})=
\sigma_{{\gamma}p \to P_{bi}^+ \to {\Upsilon(1S)}p}(\sqrt{s^*},\Gamma_{bi}), \,\,\,i=1, 2, 3.
\end{equation}

Accounting for the fact that the most of the narrow $P_{bi}^+$ and $P_{bi}^0$ resonances ($i=1$, 2, 3),
having vacuum total decay widths in their rest frames of
9.8, 20.6 and 6.4 MeV [9], respectively, decay to ${\Upsilon(1S)}p$ and ${\Upsilon(1S)}n$
outside of the considered target nuclei as well as the results presented both in Ref. [64] and above by
Eqs. (26), (27), (50), (63), we can get the following expression for the $\Upsilon(1S)$
inclusive differential cross section (momentum distribution) stemming from the production and
decay of pentaquark resonances $P_{bi}^+$ and $P_{bi}^0$ in ${\gamma}A$ reactions:
%formula(64)
\begin{equation}
\frac{d\sigma_{{\gamma}A\to {\Upsilon(1S)}X}^{({\rm sec,i})}(p_{\gamma},p_{\Upsilon(1S)})}{dp_{\Upsilon(1S)}}=
2{\pi}I_{V}[A,\sigma^{\rm in}_{{P_{b}}N}]
\int\limits_{\cos10^{\circ}}^{1}d\cos{{\theta_{\Upsilon(1S)}}}
\left<\frac{d\sigma_{{\gamma}p \to P_{bi}^+ \to {\Upsilon(1S)}p}(p_{\gamma},
p_{\Upsilon(1S)},\theta_{\Upsilon(1S)})}{dp_{\Upsilon(1S)}d{\bf \Omega}_{\Upsilon(1S)}}\right>_A,
\end{equation}
where
$\left<\frac{d\sigma_{{\gamma}p \to P_{bi}^+ \to {\Upsilon(1S)}p}(p_{\gamma},
p_{\Upsilon(1S)},\theta_{\Upsilon(1S)})}{dp_{\Upsilon(1S)}d{\bf \Omega}_{\Upsilon(1S)}}\right>_A$
($i=$1, 2, 3) is the off-shell inclusive differential cross section for the production of $\Upsilon(1S)$ mesons
with momentum ${\bf p}_{\Upsilon(1S)}$ in production/decay chain ${\gamma}p \to P_{bi}^+ \to {\Upsilon(1S)}p$,
averaged over the Fermi motion and binding energy of the protons in the nucleus.
It can be expressed by Eqs. (54)--(60) from Ref. [64], in which one needs to make the
substitutions: $J/\psi \to \Upsilon(1S)$, $P^+_{ci} \to P^+_{bi}$, $\Gamma^+_{ci} \to \Gamma_{bi}$
$E^+_{ci} \to E^+_{bi}$, ${\bf p}^+_{ci} \to {\bf p}^+_{bi}$ and $\gamma^+_{ci} \to \gamma^+_{bi}$.
For better readability of this paper, we do not give these expressions here.
The quantity $I_{V}[A,\sigma^{\rm in}_{P_{b}N}]$ in Eq. (64) is defined above by Eq. (27), in which one
needs to make the substitution $\sigma \to \sigma^{\rm in}_{P_{b}N}$.
Here, $\sigma^{\rm in}_{P_{b}N}$ is the $P_{b}$--nucleon inelastic total cross section.
For this cross section we will use in our calculations the same value of 33.5 mb as that adopted in Ref. [64]
for the inelastic cross section $\sigma^{\rm in}_{P_{c}N}$.

Before closing this subsection, we determine the free space resonant $\Upsilon(1S)$ energy distributions
$d\sigma_{{\gamma}p \to P_{bi}^+ \to {\Upsilon(1S)}p}[\sqrt{s},p_{\Upsilon(1S)}]/ dE_{\Upsilon(1S)}$
from the production/decay chains (53)/(55), proceeding on the target proton at rest,
in addition to that from the background ${\gamma}p \to {\Upsilon(1S)}p$ reaction (cf. Eq. (51)).
The energy-momentum conservation in these chains leads to the conclusion that the kinematical characteristics
of $\Upsilon(1S)$ mesons produced in them and in this reaction are the same at given incident photon energy.
Using Eq. (62) from Ref. [64] with accounting for the substitutions mentioned above, which we need to make in the respective formulas when going from the $P_c$ production and decay to the $P_b$ ones, we represent the energy distributions
$d\sigma_{{\gamma}p \to P_{bi}^+ \to {\Upsilon(1S)}{p}}[\sqrt{s},p_{\Upsilon(1S)}]/ dE_{\Upsilon(1S)}$
in the following form:
%formula(65)
\begin{equation}
\frac{d\sigma_{{\gamma}p \to P_{bi}^+ \to {\Upsilon(1S)}{p}}[\sqrt{s},p_{\Upsilon(1S)}]}{dE_{\Upsilon(1S)}}=
\sigma_{{\gamma}p \to P_{bi}^+}(\sqrt{s},\Gamma_{bi})\theta[\sqrt{s}-(m_{\Upsilon(1S)}+m_{p})]\times
\end{equation}
$$
\times
\left(\frac{\sqrt{s}}{2p_{\gamma}p^{*}_{\Upsilon(1S)}}\right)
Br[P_{bi}^+ \to {\Upsilon(1S)}p]~{\rm for}
~E^{(2)}_{\Upsilon(1S)}(0^{\circ}) \le E_{\Upsilon(1S)} \le E^{(1)}_{\Upsilon(1S)}(0^{\circ}).
$$
Eq. (65) shows that the free space $\Upsilon(1S)$ energy spectrum, which arises from the
production/decay chains (53)/(55), exhibits a completely flat behavior within the
allowed energy range (18).
%%%%%%%%%%%%%%%%%%%%%%%%%%%%%%%%%%%%%%%%%%%%%%%%%%%%%%%%%%%
\begin{figure}[h!]
\begin{center}
\includegraphics[width=16.0cm]{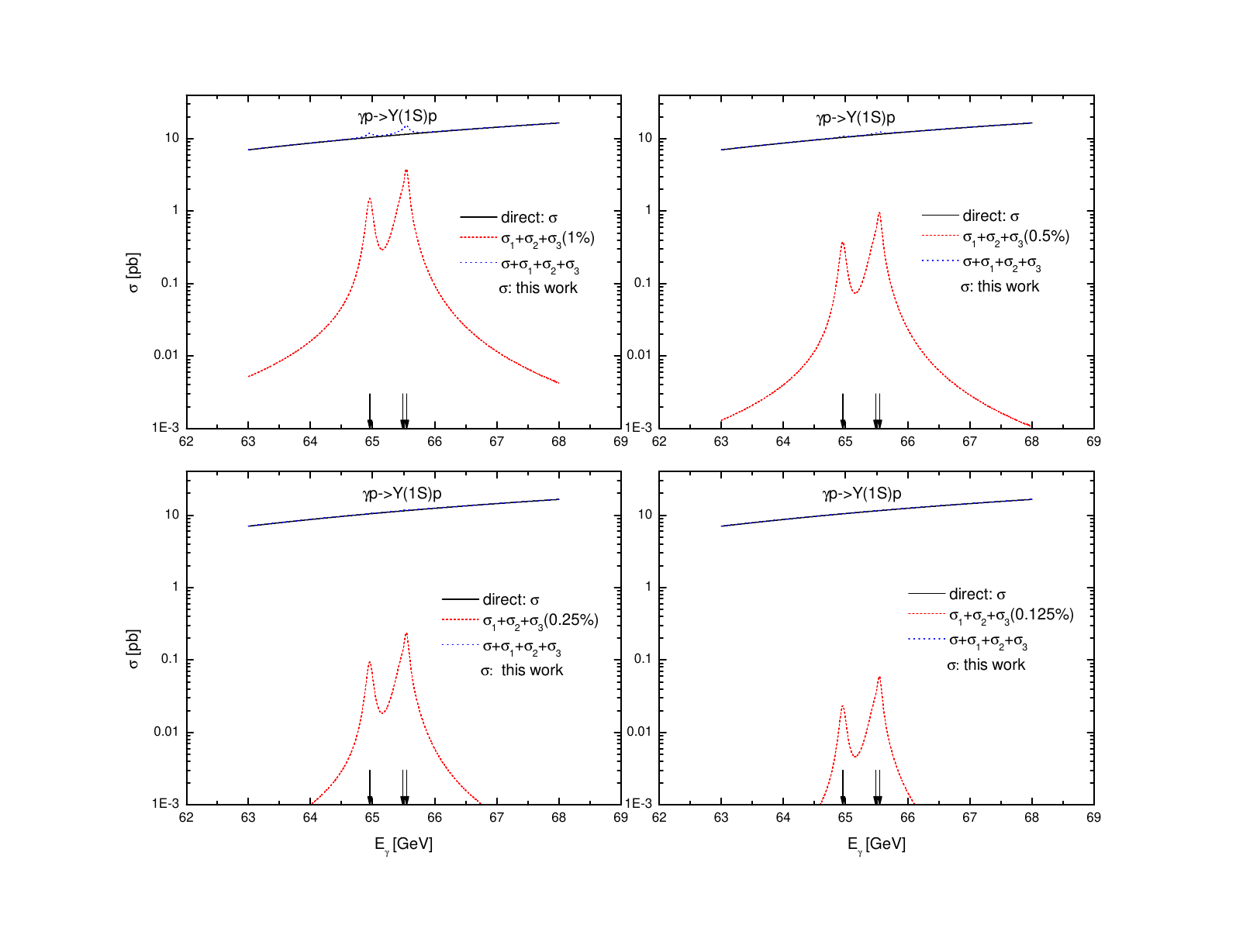}
\vspace*{-2mm} \caption{(Color online.) The non-resonant total cross section $\sigma$ for the reaction
${\gamma}p \to {\Upsilon(1S)}p$ (black solid curve), calculated on the basis of Eq. (40).
Incoherent sum (blue dotted curve) of it and the sum
$\sigma_1$+$\sigma_2$+$\sigma_3$ (red short-dashed curve) of the total cross sections $\sigma_1$,
$\sigma_2$ and $\sigma_3$ for the
resonant $\Upsilon(1S)$ production in the processes ${\gamma}p \to P_{b}^+(11080) \to {\Upsilon(1S)}p$,
${\gamma}p \to P_{b}^+(11125) \to {\Upsilon(1S)}p$
and ${\gamma}p \to P_{b}^+(11130) \to {\Upsilon(1S)}p$, calculated in line with Eq. (60)
assuming that the resonances
$P_{b}^+(11080)$, $P_{b}^+(11125)$ and $P_{b}^+(11130)$ with the spin-parity quantum
numbers $J^P=(1/2)^-$, $J^P=(1/2)^-$ and $J^P=(3/2)^-$
decay to the ${\Upsilon(1S)}p$ channels with all three individual branching fractions
1, 0.5, 0.25 and 0.125\% (respectively, upper left, upper right,
lower left and lower right panels), as functions of laboratory photon energy $E_{\gamma}$.
Three arrows indicate resonant energies
$E^{\rm R1}_{\gamma}=64.952$ GeV, $E^{\rm R2}_{\gamma}=65.484$ GeV
and $E^{\rm R3}_{\gamma}=65.544$ GeV.}
\label{void}
\end{center}
\end{figure}
%%%%%%%%%%%%%%%%%%%%%%%%%%%%%%%%%%%%%%%%%%%%%%%%%%%%%%%%%%%%%%%%%%%%%%%%%%%%%%%%%%%%%%%%%%%%%
\begin{figure}[h!]
\begin{center}
\includegraphics[width=16.0cm]{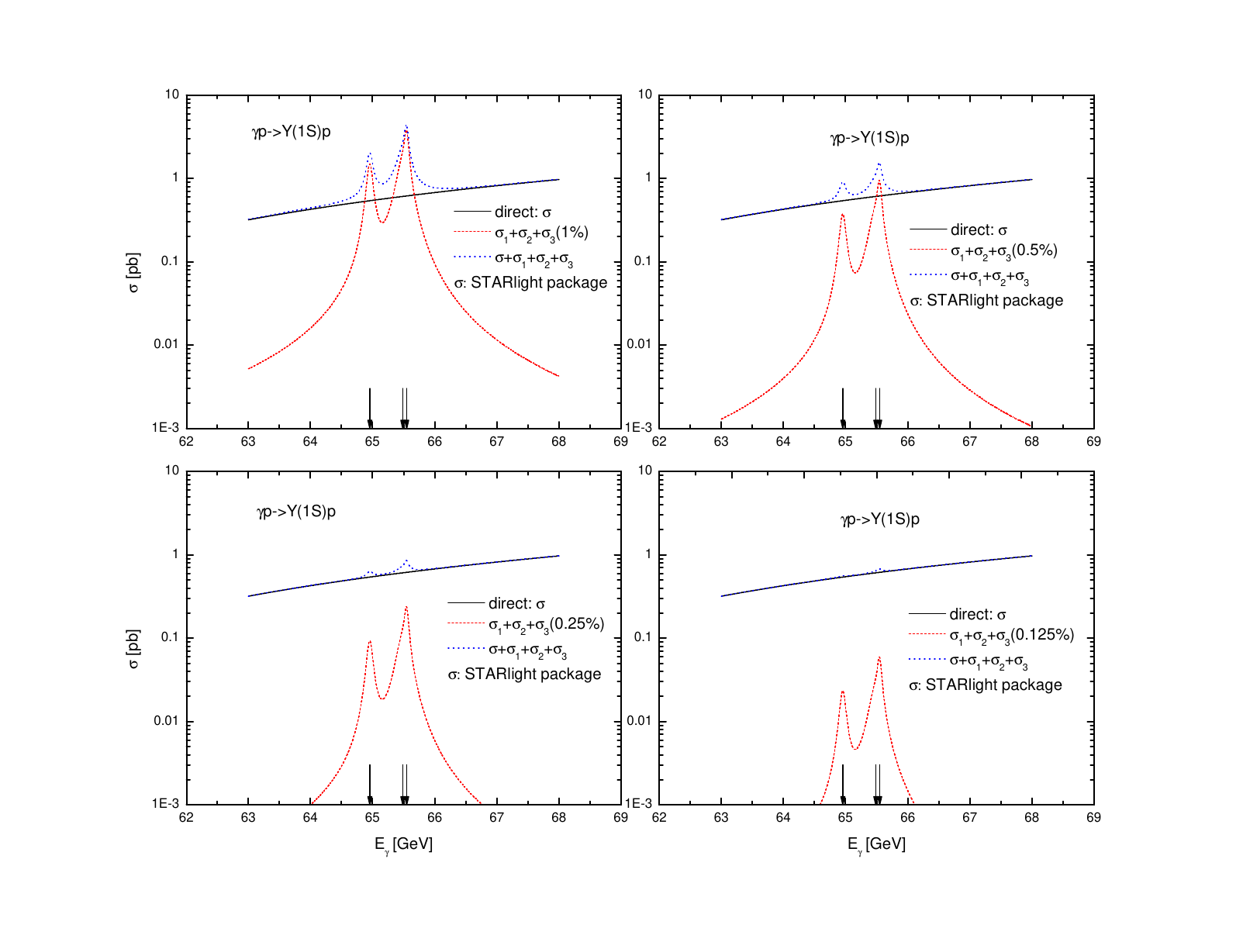}
\vspace*{-2mm} \caption{(Color online.) The same as in Fig. 5, but the non-resonant total cross section $\sigma$
for the reaction ${\gamma}p \to {\Upsilon(1S)}p$ is calculated on the basis of Eq. (45).}
\label{void}
\end{center}
\end{figure}
%%%%%%%%%%%%%%%%%%%%%%%%%%%%%%%%%%%%%%%%%%%%%%%%%%%%%%%%%%%

\section*{3. Results and discussion}

  The free direct non-resonant $\Upsilon(1S)$ production total cross section (40) on the target proton being at rest
(black solid curves), the total cross section for the resonant $\Upsilon(1S)$ production in the production/decay chains (53)/(55), determined based on Eq. (60) for the considered
branching ratios $Br[P^+_{bi} \to {\Upsilon(1S)}p]=$ 0.125, 0.25, 0.5 and 1\% for all three
$P^+_{bi}$ ($i=$1, 2, 3) states (red short-dashed curves) and the combined (non-resonant plus resonant)
$\Upsilon(1S)$ production total cross section (blue dotted curves) are depicted in Fig. 5 as functions of photon
energy. Fig. 6 shows the same as that presented in Fig. 5, but here the non-resonant total cross section $\sigma$
for the reaction ${\gamma}p \to {\Upsilon(1S)}p$ is calculated on the basis of the parametrization (45), used
in the STARlight Monte Carlo simulation program, to see the sensitivity of the combined
$\Upsilon(1S)$ production total cross section to the possible background contribution.
It can be seen from these figures that the $P^+_b(11080)$ state as well as $P^+_b(11125)$ and $P^+_b(11130)$
resonances appear, correspondingly, as clear narrow independent peak as well as one distinct wide peak
(due to low distance between their centroids--60 MeV)
at resonant photon energies $E_{\gamma}=$ 64.952 GeV as well as at $E_{\gamma}$ $\approx$ 65.50 GeV
in the combined cross section, only if this contribution is chosen in the form of Eq. (45) and
$Br[P^+_{bi} \to {\Upsilon(1S)}p]=$~1\% for all three $P^+_{bi}$ states.
In this case, at laboratory photon energies around the peak energies the resonant contributions are
significantly larger than the non-resonant ones of about 0.6 pb and the peak values of the combined cross section
reach a well measurable values $\sim$ 2--4 pb. Therefore, the background reaction ${\gamma}p \to {\Upsilon}p$ will
not influence the direct observation of the hidden-bottom pentaquark production at these energies and in this case.
If $Br[P^+_{bi} \to {\Upsilon(1S)}p] \le$ 0.5\% ($i=$1, 2, 3), then the resonant $\Upsilon(1S)$ yields are comparable to (or are less than) the non-resonant ones in the resonance regions. As a result, the considered pentaquark states do not manifest itself as clear peaks in the combined cross section.
This means that will be difficult to measure these states in $\Upsilon(1S)$ total production cross section on a proton target in the photon-induced reactions in such cases.
From Fig. 5 we see yet that the resonant $\Upsilon(1S)$ production cross section is small compared to the non-resonant
contribution at all considered photon energies for four employed values for the branching fractions
$Br[P^+_{bi} \to {\Upsilon(1S)}p]$ in the case when the background contribution is chosen in the form of Eq. (40).
In this case, the combined total cross section of the reaction ${\gamma}p \to {\Upsilon(1S)}p$ has no clear peak structures, corresponding to the $P^+_{bi}$ pentaquark states, and it is practically not distinguished from that for
the background reaction. In line with the preceding, one may  expect that the $P^+_{bi}$ signals could be well distinguished from the background reaction via the detailed scan of the $\Upsilon(1S)$ total production cross section on a proton target in the near-threshold incident photon energy regions around the resonant energies of 64.952 and $\approx$ 65.50 GeV in the future photoproduction experiments at electron-ion colliders, only if the background contribution is of the order of that given by the parametrization (45), used in the STARlight package, and if the branching ratios
$Br[P^+_{bi} \to {\Upsilon(1S)}p]$ $\sim$ 1\%.
To experimentally observe a two-peak structure, corresponding to the $P^+_b(11080)$ and $P^+_b(11125)$,
$P^+_b(11130)$ states, in the combined total cross section of the reaction
${\gamma}p \to {\Upsilon(1S)}p$, it is sufficient to have a photon energy resolution and energy binning
of the order of 20--30 MeV. Thus, the c.m. energy ranges
$M_{bi}-{\Gamma_{bi}}/2 < \sqrt{s} < M_{bi}+{\Gamma_{bi}}/2$ ($i=$1, 2) correspond to the laboratory
photon energy regions of 64.894 GeV $< E_{\gamma} <$ 65.010 GeV
and 65.362 GeV $< E_{\gamma} <$ 65.607 GeV, i.e. ${\Delta}E_{\gamma}=$ 116
and 245 MeV for $P^+_{b}(11080)$ and $P^+_{b}(11125)$, respectively.
This means that to resolve the two peaks in the upper left panel of Fig. 6, a photon energy resolution
and energy bin size of the order of 20--30 MeV are required. One may hope that this requirement will be
satisfied in the $\Upsilon(1S)$ photoproduction experiments at the future electron-ion colliders ( cf. [70]).
Since the ${\gamma}p$ and $ep$ reactions are very similar, the hidden-bottom pentaquarks $P^+_{bi}$ can also be searched for on them directly via the $ep$ scattering process. To motivate the searching for and studying of these pentaquark states in this process, it is highly desirable to evaluate their production rates in it. For this purpose, we
translate the $\Upsilon(1S)$ photoproduction total cross section predictions, reported above, into
the expected yields of the $P^+_{bi}$ signals from the reactions $ep \to eP^+_{bi} \to e{\Upsilon(1S)}p$, $\Upsilon(1S) \to l^+l^-$
\footnote{$^)$The lepton pairs $l^+l^-$ denote both ${\mu^+}{\mu^-}$ and $e^+e^-$.}$^)$
at least in the case, corresponding to the upper left panel of Fig. 6.
According to current point of view [65, 66], one year of running of EicC (China) would deliver
an integrated luminosity of 50 fb$^{-1}$. For the pentaquark yields (for the total numbers of the $P^+_b(11080)$ and $P^+_b(11125)$, $P^+_b(11130)$ events) estimates in a one-year run,
one needs to multiply the above luminosity by the $P^+_{b}(11080)$ and $P^+_b(11125)$, $P^+_b(11130)$
production cross sections of 1.4 and 3.4 pb at resonant photon energies of 64.952 and 65.50 GeV
and by the detection efficiency and by the appropriate branching ratio
$Br[{\Upsilon(1S)} \to l^+l^-]$ $\approx$ 5\% as well as to account for the fact that approximately two orders of magnitude smaller cross sections are anticipated for the electroproduction compared to the above photon production [62].
With a realistic 30\% detection efficiency [66], we estimate about of 10 and 25 events
per year for the $P^+_{b}(11080)$ and $P^+_{b}(11125)$, $P^+_{b}(11130)$ signals, respectively
\footnote{$^)$It should be mentioned that these numbers of events are in good agreement with the production
rates of two narrow $P_b$ states at EicC, presented in Table 2.3 of Ref. [66].}$^)$
.
The accumulated luminosity with one-year run can be reached of about 100 fb$^{-1}$ at the EIC (US)
facility [70]. Under this circumstance and in line with the estimates given above, the expected numbers of the
respective $P^+_{bi}$ events of around 20 and 50 are achievable with one-year of EIC running.
For elastic (background) ${\Upsilon(1S)}p$ production in the reaction ${\gamma}p \to {\Upsilon(1S)}p$ we have
$\sigma \approx 0.6$ pb at considered photon energies. This leads to $\approx$ 4 events per year. We see that
the elastic background is sufficiently small compared to the resonant yields.
Therefore, the observation of hidden-bottom pentaquarks is still promising at EicC and EIC also with the
electromagnetic $ep$ reactions.
It is obvious that to see the $P^+_{bi}$ pentaquarks  experimentally in photoproduction experiments in the cases,
when more "softer" constraints than those given above are imposed on the possible background contribution and on the branching ratios $Br[P^+_{bi} \to {\Upsilon(1S)}p]$, one needs to consider such observable, which is appreciably
sensitive to the $P^+_{b}$ signal in some region of the available phase space.
For instance, the large $t$ region of the differential cross section $d\sigma/dt$,
where the diffractive $t$-dependence of the background $\Upsilon(1S)$ meson production
is suppressed while its resonant production is expected to be rather flat (cf. Refs. [112, 113]).
%%%%%%%%%%%%%%%%%%%%%%%%%%%%%%%%%%%%%%%%%%%%%%%%%%%%%%%%%%%
\begin{figure}[!h]
\begin{center}
\includegraphics[width=16.0cm]{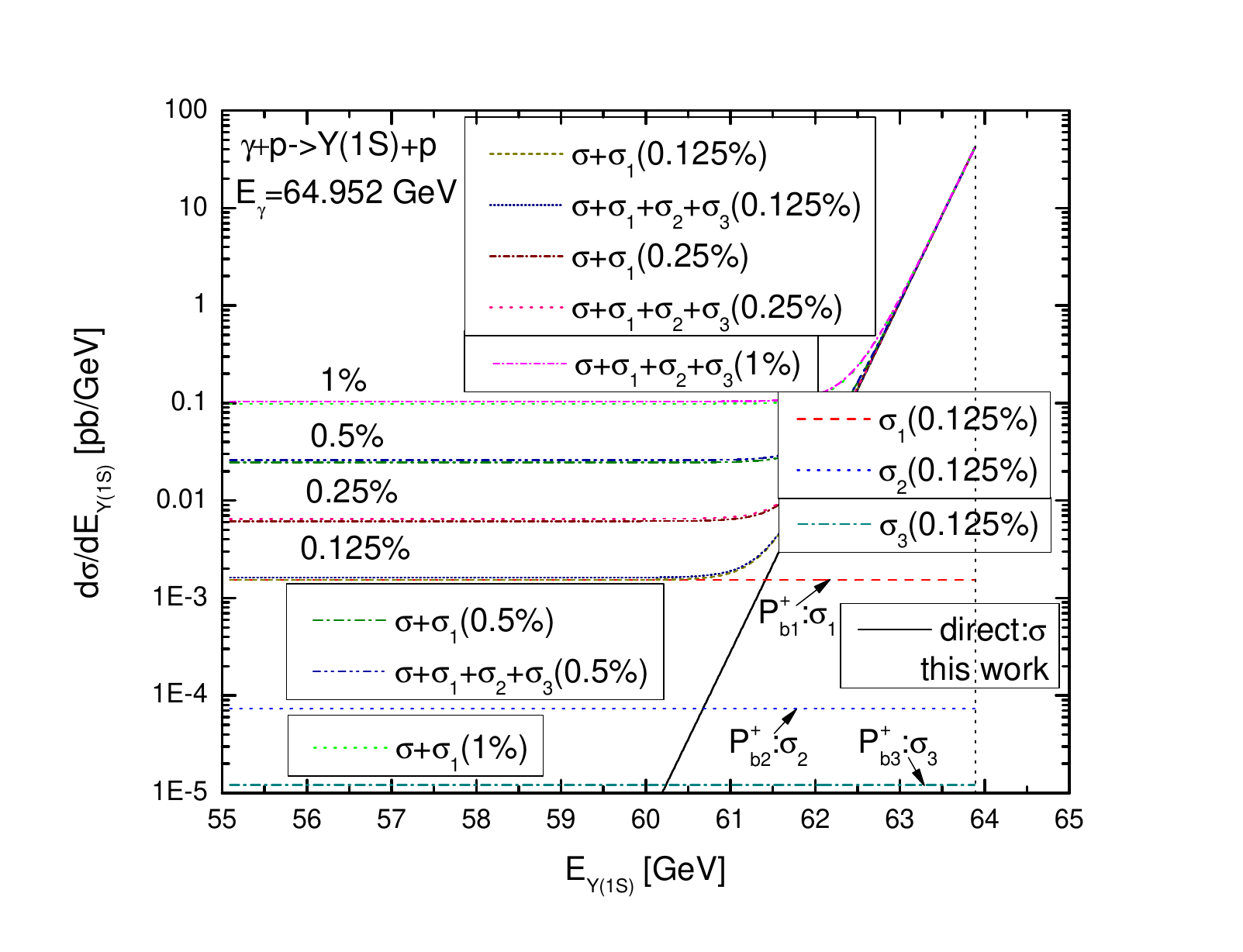}
\vspace*{-2mm} \caption{(Color online.) The direct non-resonant $\Upsilon(1S)$ energy distribution in the free space
elementary process ${\gamma}p \to {\Upsilon(1S)}p$,
calculated in line with Eqs. (40), (51) at initial photon resonant energy of 64.952 GeV
in the laboratory system (black solid curve). The resonant $\Upsilon(1S)$ energy distributions
in the two-step processes
${\gamma}p \to P_{b}^+(11080) \to {\Upsilon(1S)}p$,
${\gamma}p \to P_{b}^+(11125) \to {\Upsilon(1S)}p$ and ${\gamma}p \to P_{b}^+(11130) \to {\Upsilon(1S)}p$,
calculated in line with Eq. (65) at the same incident photon energy of 64.952 GeV
assuming that the resonances $P_{b}^+(11080)$, $P_{b}^+(11125)$, $P_{b}^+(11130)$
with the spin-parity assignments $J^P=(1/2)^-$, $J^P=(1/2)^-$, $J^P=(3/2)^-$, correspondingly,
all decay to the ${\Upsilon(1S)}p$ mode
with branching fractions of 0.125\% (respectively, red dashed, blue dotted, dark cyan dashed-doted curves).
Incoherent sum of the direct non-resonant $\Upsilon(1S)$ energy distribution and resonant ones, calculated supposing
that the resonances $P_{b}^+(11080)$ as well as $P_{b}^+(11080)$, $P_{b}^+(11125)$, $P_{b}^+(11130)$
with the same spin-parity combinations all decay to the ${\Upsilon(1S)}p$ with branching
fractions 0.125, 0.25, 0.5, 1\% (respectively, dark yellow short-dashed,
wine short-dashed-dotted, olive dashed-dotted, green dotted as well as navy short-dotted, pink dotted, royal dashed-dotted-dotted, magenta  short-dashed-dotted curves), all as functions of the total $\Upsilon(1S)$ energy
$E_{\Upsilon(1S)}$ in the laboratory system.
The vertical dotted line indicates the upper $\Upsilon(1S)$ allowed energy in this system (63.893 GeV)
for the considered direct non-resonant and resonant $\Upsilon(1S)$ production on a free target proton at rest at
given initial photon resonant energy of 64.952 GeV.}
\label{void}
\end{center}
\end{figure}
%%%%%%%%%%%%%%%%%%%%%%%%%%%%%%%%%%%%%%%%%%%%%%%%%%%%%%%%%%%
\begin{figure}[!h]
\begin{center}
\includegraphics[width=16.0cm]{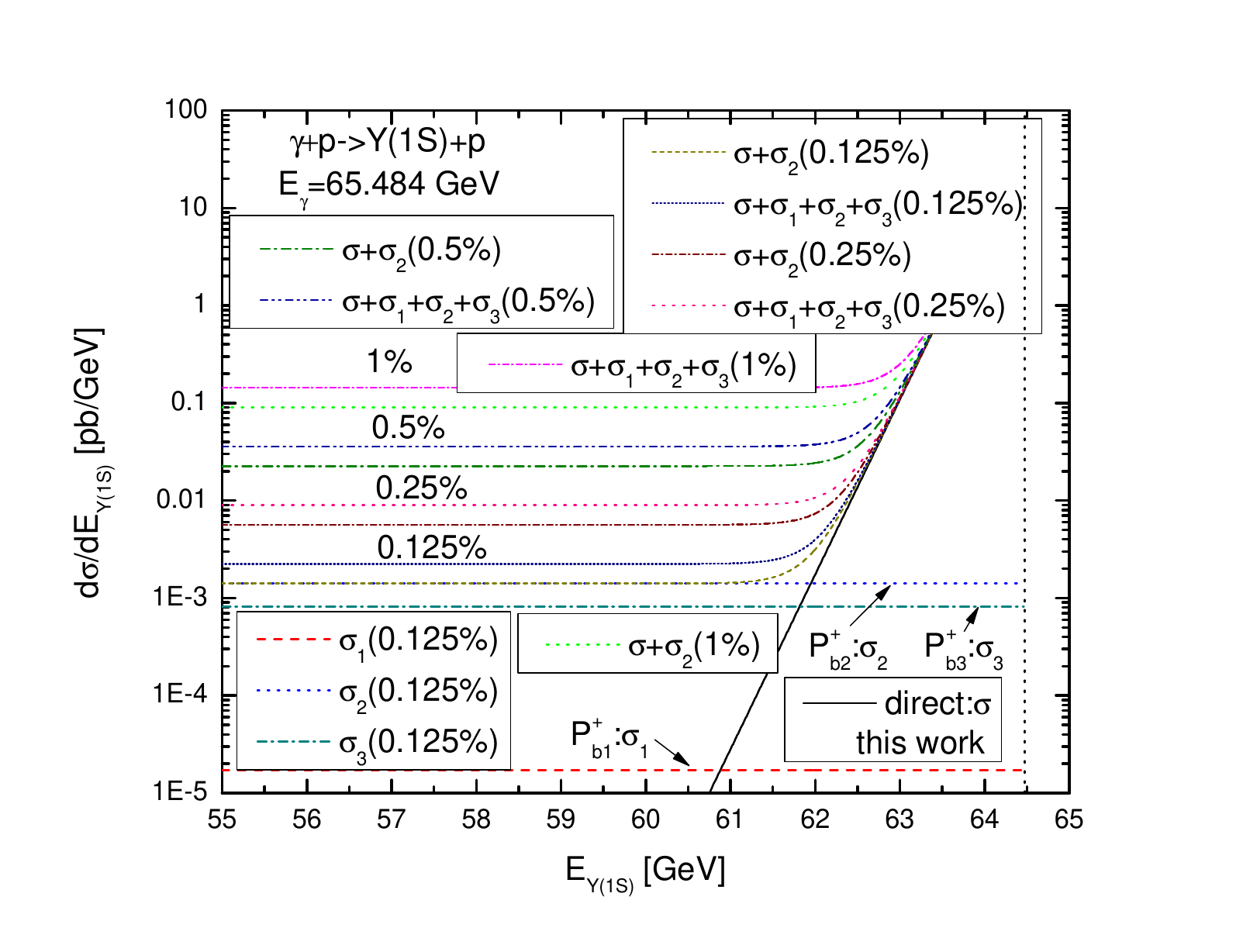}
\vspace*{-2mm} \caption{(Color online.) The same as in Fig. 7, but for initial photon resonant energy of 65.484 GeV
in the laboratory system. And here, incoherent sum of the direct non-resonant $\Upsilon(1S)$ energy distribution and resonant ones is calculated for the resonance $P_{b}^+(11125)$ as well as for the $P_{b}^+(11080)$, $P_{b}^+(11125)$, $P_{b}^+(11130)$ states.}
\label{void}
\end{center}
\end{figure}
%%%%%%%%%%%%%%%%%%%%%%%%%%%%%%%%%%%%%%%%%%%%%%%%%%%%%%%%%%%
\begin{figure}[!h]
\begin{center}
\includegraphics[width=16.0cm]{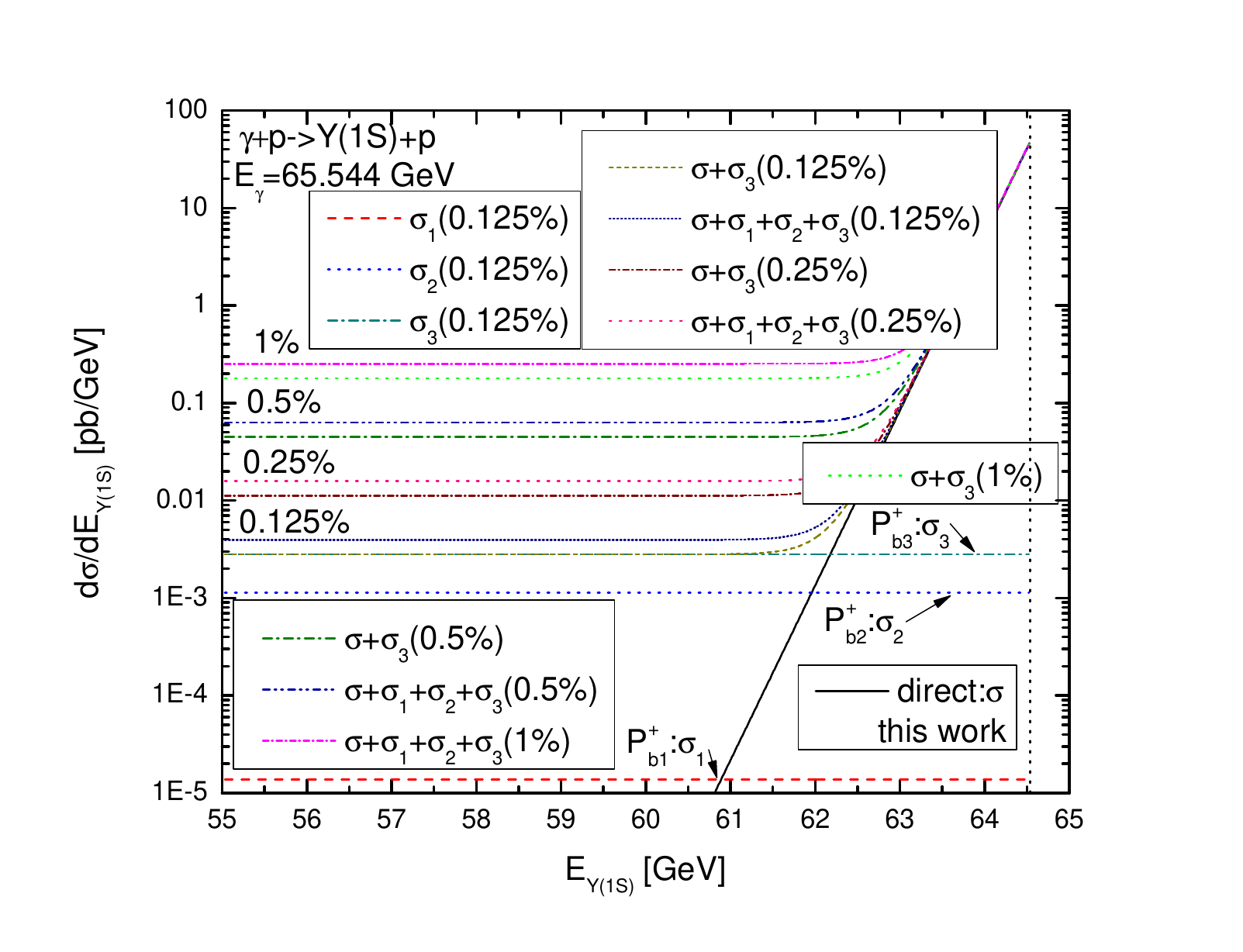}
\vspace*{-2mm} \caption{(Color online.)
The same as in Fig. 7, but for initial photon resonant energy of 65.544 GeV
in the laboratory system. And here, incoherent sum of the direct non-resonant $\Upsilon(1S)$ energy distribution and resonant ones is calculated for the resonance $P_{b}^+(11130)$ as well as for the $P_{b}^+(11080)$, $P_{b}^+(11125)$, $P_{b}^+(11130)$ states.}
\label{void}
\end{center}
\end{figure}
%%%%%%%%%%%%%%%%%%%%%%%%%%%%%%%%%%%%%%%%%%%%%%%%%%%%%%%%%%%%%%%%%%%%%%%%%%%%%%%%%%%%%%%%%%%%%
%%%%%%%%%%%%%%%%%%%%%%%%%%%%%%%%%%%%%%%%%%%%%%%%%%%%%%%%%%%
\begin{figure}[h!]
\begin{center}
\includegraphics[width=16.0cm]{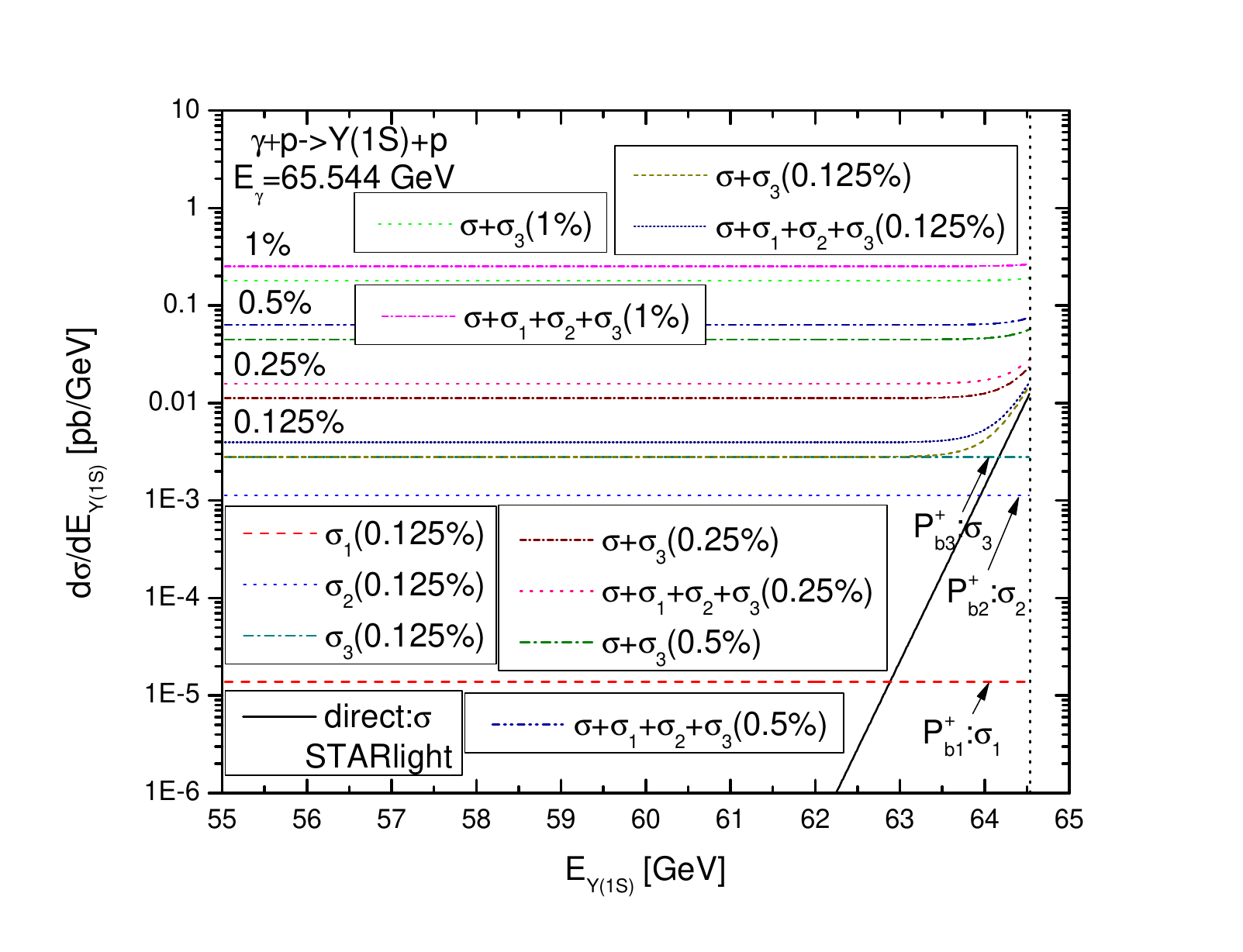}
\vspace*{-2mm} \caption{(Color online.) The same as in Fig. 9, but the non-resonant total cross section $\sigma$
for the reaction ${\gamma}p \to {\Upsilon(1S)}p$ is calculated on the basis of Eq. (45).}
\label{void}
\end{center}
\end{figure}
%%%%%%%%%%%%%%%%%%%%%%%%%%%%%%%%%%%%%%%%%%%%%%%%%%%%%%%%%%%%%%%%%%%%%%%%%%%%%%%%%%%%%%%%%%%%%

  In line with the above-mentioned, we consider now the $\Upsilon(1S)$ energy distribution
from the ${\gamma}p \to {\Upsilon(1S)}p$ elementary reaction. The direct non-resonant $\Upsilon(1S)$
energy distribution from this reaction and the resonant ones from the production/decay chains (53)/(55),
proceeding on the free target proton at rest, were calculated in line with Eqs. (40), (51) and (65), respectively,
for incident photon resonant energies of 64.952, 65.484 and 65.544 GeV.
The resonant $\Upsilon(1S)$ energy distributions were determined for the adopted spin-parity assignments
of the $P^+_{b}(11080)$, $P^+_{b}(11125)$, $P^+_{b}(11130)$ resonances for branching fractions
$Br[P^+_{bi} \to {\Upsilon(1S)}p]=$~0.125\% for all three states.
These dependencies, together with the incoherent sum of the non-resonant $J/\psi$ energy distribution
and resonant ones, calculated assuming that all the resonances
$P^+_{b}(11080)$ and $P^+_{bi}$ ($i=1$, 2, 3), $P^+_{b}(11125)$ and $P^+_{bi}$ ($i=1$, 2, 3),
$P^+_{b}(11130)$ and $P^+_{bi}$ ($i=1$, 2, 3) decay to the ${\Upsilon(1S)}p$ mode
with four adopted options for the branching ratios $Br[P^+_{bi} \to {\Upsilon(1S)}p]$,
as functions of the $\Upsilon(1S)$ total energy $E_{\Upsilon(1S)}$ are shown, respectively, in Figs. 7, 8, 9.
The same as that shown in Fig. 9, but calculated for the non-resonant total cross section $\sigma$
for the reaction ${\gamma}p \to {\Upsilon(1S)}p$ taken in the form of Eq. (45) used in the STARlight program
package, is presented in Fig. 10 to test the sensitivity of the combined $\Upsilon(1S)$ energy distribution to the
choice of the background contribution.
One can see from these figures that while the resonant $\Upsilon(1S)$ production differential
cross sections show a flat behavior at allowed energies $E_{\Upsilon(1S)}$,
the non-resonant differential cross section drops quickly as the energy $E_{\Upsilon(1S)}$ decreases.
At initial photon resonant energies of 64.952, 65.484 and 65.544 GeV of interest and in the case when
the direct non-resonant $\Upsilon(1S)$ energy distribution was calculated in line with Eqs. (40), (51)
(Figs. 7--9) its strength is significantly larger than those of the resonant $\Upsilon(1S)$ production
cross sections, determined for the value of the branching ratios $Br[P^+_{bi} \to {\Upsilon(1S)}p]=$~0.125\%,
for "high" allowed $\Upsilon(1S)$ total energies greater than $\approx$~62 GeV.
Whereas at "low" $\Upsilon(1S)$ total energies (below 62 GeV)
the contribution from the resonance, decaying to the ${\Upsilon(1S)}p$ final state with the branching ratio of 0.125\%,
with the centroid, respectively, at photon energies of 64.952, 65.484 and 65.544 GeV, is much larger than that from the direct non-resonant process ${\gamma}p \to {\Upsilon(1S)}p$.
When the non-resonant total cross section $\sigma$ for the reaction ${\gamma}p \to {\Upsilon(1S)}p$ is taken in the form of Eq. (45) used in the STARlight package, analogous $\Upsilon(1S)$ total energy corresponding to the initial photon
energy of $E_{\gamma}=$~65.544 GeV is somewhat greater, and it is $\approx$~64 GeV (see Fig. 10).
Thus, for example, in the first and second cases (what concerns the choices for the background contribution
considered), for the $\Upsilon(1S)$ mesons with total energies about of 61 as well as 63 GeV
their resonant production cross section, calculated for the branching fraction of 0.125\% of its decay to the
${\Upsilon(1S)}p$ mode, is enhanced compared to the non-resonant one by essential factors
about of one and two order of magnitude at incident photon energies of 64.952 and 65.484, 65.544 GeV, respectively,
as well as by factor about of two order of magnitude at photon energy of 65.544 GeV.
Moreover, this contribution is also noticeably larger than those, appearing from the decays of another two hidden-bottom pentaquarks to the ${\Upsilon(1S)}p$ with the branching ratios
$Br[P^+_{bi} \to {\Upsilon(1S)}p]=$~0.125\% at the above-mentioned "low" $\Upsilon(1S)$ total energies.
As a result, at each considered resonant photon energy the $\Upsilon(1S)$ meson combined energy distribution,
arising from its direct production and from the decay of the intermediate pentaquark resonance,
located at this energy, to the ${\Upsilon(1S)}p$ final state reveals here a clear sensitivity to the
adopted options for the branching ratio of this decay.
Thus, for instance, for the $\Upsilon(1S)$ mesons with total energy of 61 GeV and for the lowest incident photon
energy of 64.952 GeV this $\Upsilon(1S)$ combined distribution is enhanced
for the values of this ratio of 0.125, 0.25, 0.5, 1\% by significant
factors of about 5, 25, 90, 350, respectively, as compared to that from the directly produced $\Upsilon(1S)$
mesons in the case when the direct non-resonant $\Upsilon(1S)$ energy distribution was calculated in line with Eqs. (40), (51) (cf. Fig. 7). And in this case for the highest initial photon resonant energy of 65.544 GeV
of interest, the analogous factors become much larger and they are of about 1.3$\cdot$10$^2$, 5.0$\cdot$10$^2$,
2.0$\cdot$10$^3$, 8.2$\cdot$10$^3$, respectively (see Fig. 9).
Similarly, for this initial photon energy, for the $\Upsilon(1S)$ mesons with total energy of 63 GeV, and
in the case when the direct $\Upsilon(1S)$ energy distribution was calculated in line with Eqs. (45), (51)
these factors are practically the same as above (cf. Figs. 9, 10).
It is important to emphasize that, as can be seen from Figs. 7--10, the above "partial"
combined energy distribution of $\Upsilon(1S)$ mesons, calculated at given initial photon resonant energy
for given values of the branching ratios $Br[P^+_{bi} \to {\Upsilon(1S)}p]$ ($i=$1, 2, 3),
is practically indistinguishable from their "total" combined energy distribution, arising from the direct and resonant $\Upsilon(1S)$ production via all production/decay processes (53)/(55), and determined at the same photon energy and the same values of these branching ratios. It is also important that these "partial" and "total" combined differential energy distributions have, contrary to the $\Upsilon(1S)$ total production cross sections (cf. Figs. 5, 6), a weak sensitivity to the choice of the background contribution at "low" $\Upsilon(1S)$ total energies smaller than $\approx$~62 GeV (compare Figs. 9 and 10). At these $\Upsilon(1S)$ energies,
the differences between the combined results, obtained by using a value of the branching fractions of the decays $P^+_{bi} \to {\Upsilon(1S)}p$ of 0.125\% and the non-resonant background, as well as between the combined results, determined by adopting the values of the branching ratios of these decays
of 0.125 and 0.25\%, 0.25 and 0.5\%, 0.5 and 1\%, are quite large and experimentally measurable
for both adopted options for the background contribution.
Further, for each initial photon resonant energy considered
the observation here of the specific hidden-bottom pentaquark will be practically
not influenced by the presence of the another two hidden-bottom pentaquarks and by the background reaction.
Since the $\Upsilon(1S)$ photoproduction differential cross sections have a small absolute values
$\sim$ 0.001--0.1 pb/GeV at "low" $\Upsilon(1S)$ total energies considered for branching fractions
$Br[P^+_{bi} \to {\Upsilon(1S)}p]$ ($i=$1, 2, 3) of 0.125--1\%, respectively,
their measurements require both high luminosities and large-acceptance detectors. Such measurements might be performed in the future at the electron-ion colliders EicC and EIC. At these colliders, the hidden-bottom pentaquarks $P^+_{bi}$ can also be searched for and studied directly through the $ep$ interaction process. To further motivate the finding and studying of these pentaquark states in the $ep$ scattering process by means of the respective $\Upsilon(1S)$ energy distribution, it is a crucial to estimate their production rates in this observable. For this purpose, one needs
to translate the $\Upsilon(1S)$ photoproduction differential cross sections, given above, into
the expected yields of the $P^+_{bi}$ signals from the reactions $ep \to eP^+_{bi} \to e{\Upsilon(1S)}p$, $\Upsilon(1S) \to l^+l^-$. Accounting for the estimates of the $P^+_{bi}$
($i=$1, 2, 3) yields given before and these $\Upsilon(1S)$ photoproduction differential cross sections at
"low" $\Upsilon(1S)$ total energies within the range of 55--62 GeV, we can evaluate the numbers of events expected in the measurement of the $\Upsilon(1S)$ mesons in this energy range for these reactions. They are about
of 0--5 and 0--10 with the one-year run of the EicC and EIC facilities, respectively. Thus, it is still optimistic that the hypothesized $P^+_{bi}$ states can also be directly observed and studied through the low-energy $\Upsilon(1S)$ meson production in the $ep$ reactions at EicC and EIC colliders. It should be noticed that most probably it would be difficult to determine the spin-parity quantum numbers of the $P^+_{bi}$ states by employing, for example, the combined
photoproduction total and integrated over production angles differential cross sections considered incoherently in the present work since that observables, based on the squared absolute values of the resonant and non-resonant production amplitudes, depend weakly on them.
Such consideration is justified because the role of the interference effects between resonant and non-resonant
contributions in the ${\gamma}p \to {\Upsilon(1S)}p$ reaction as well as between different resonance states in the
observables considered is expected to be insignificant due to the fact that the $t$-channel $\Upsilon(1S)$ production
contributes mainly to the forward angles in the c.m. frame, while the $s$-channel resonances contribute in a full
solid angle [62]. A way of determining the $P^+_{bi}$ quantum numbers should be based on the model, which
accounts for the interference effects via adding the resonant and non-resonant production amplitudes coherently.
The results of such procedure depend, in particular, on these numbers. To probe them, another observables like
angular distributions [114, 115] or polarization observables (single and double) [109, 116] are needed.
Their consideration is beyond the scope of the present paper. But, one may hope that it will be performed in the future work when the appropriate ${\gamma}p$ (and $ep$) data will appear to control that interference.
\begin{figure}[!h]
\begin{center}
\includegraphics[width=16.0cm]{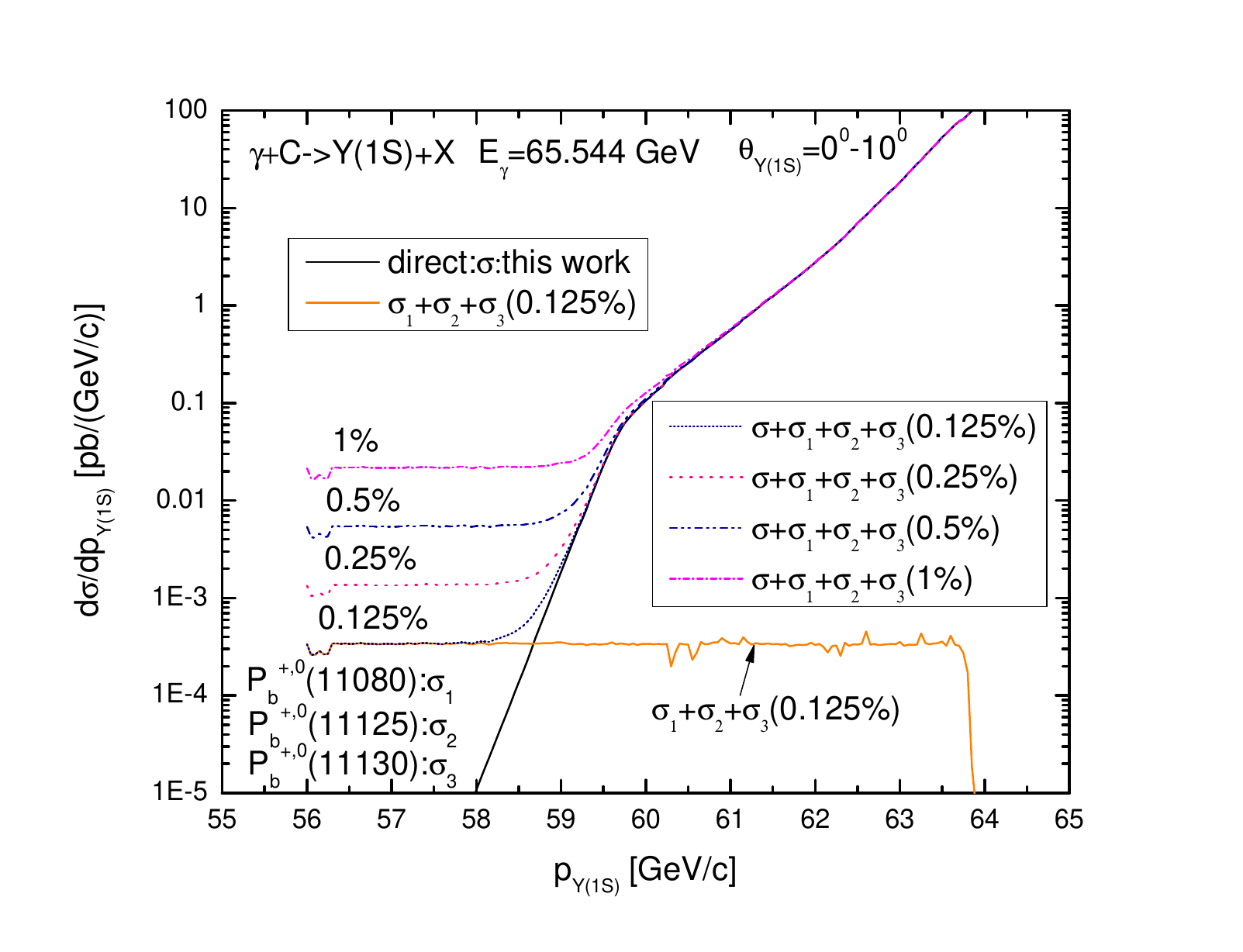}
\vspace*{-2mm} \caption{(Color online.) The direct non-resonant momentum distribution
of $\Upsilon(1S)$ mesons, produced in the reaction ${\gamma}{\rm ^{12}C} \to {\Upsilon(1S)}X$
in the laboratory polar angular range of 0$^{\circ}$--10$^{\circ}$ and calculated in line with Eqs. (40), (50)
at initial photon resonant energy of 65.544 GeV in the laboratory system (black solid curve).
Incoherent sum of the resonant momentum distributions of $\Upsilon(1S)$ mesons, produced
in the two-step processes ${\gamma}p(n) \to P_{b}^+(11080)(P_{b}^0(11080)) \to {\Upsilon(1S)}p(n)$,
${\gamma}p(n) \to P_{b}^+(11125)(P_{b}^0(11125)) \to {\Upsilon(1S)}p(n)$ and
${\gamma}p(n) \to P_{b}^+(11130)(P_{b}^0(11130)) \to {\Upsilon(1S)}p(n)$ and
calculated in line with Eq. (64) at the same incident photon energy of 65.544 GeV
assuming that the resonances $P_{b}^{+,0}(11080)$, $P_{b}^{+,0}(11125)$
and $P_{b}^{+,0}(11130)$ with the spin-parity assignments
$J^P=(1/2)^-$, $J^P=(1/2)^-$ and $J^P=(3/2)^-$, correspondingly,
all decay to the ${\Upsilon(1S)}p(n)$ modes
with branching ratios of 0.125\% (orange solid curve).
Incoherent sum of the direct non-resonant $\Upsilon(1S)$ momentum distribution and resonant ones,
calculated supposing that the resonances
$P_{b}^{+,0}(11080)$, $P_{b}^{+,0}(11125)$, $P_{b}^{+,0}(11130)$
with the same spin-parity combinations all decay to the ${\Upsilon(1S)}p(n)$ with branching
fractions 0.125, 0.25, 0.5 and 1\% (respectively, navy short-dotted, pink dotted, royal dashed-dotted-dotted and magenta short-dashed-dotted curves), all as functions of the $\Upsilon(1S)$ momentum $p_{\Upsilon(1S)}$ in the laboratory system.}
\label{void}
\end{center}
\end{figure}
%%%%%%%%%%%%%%%%%%%%%%%%%%%%%%%%%%%%%%%%%%%%%%%%%%%%%%%%%%%
\begin{figure}[!h]
\begin{center}
\includegraphics[width=16.0cm]{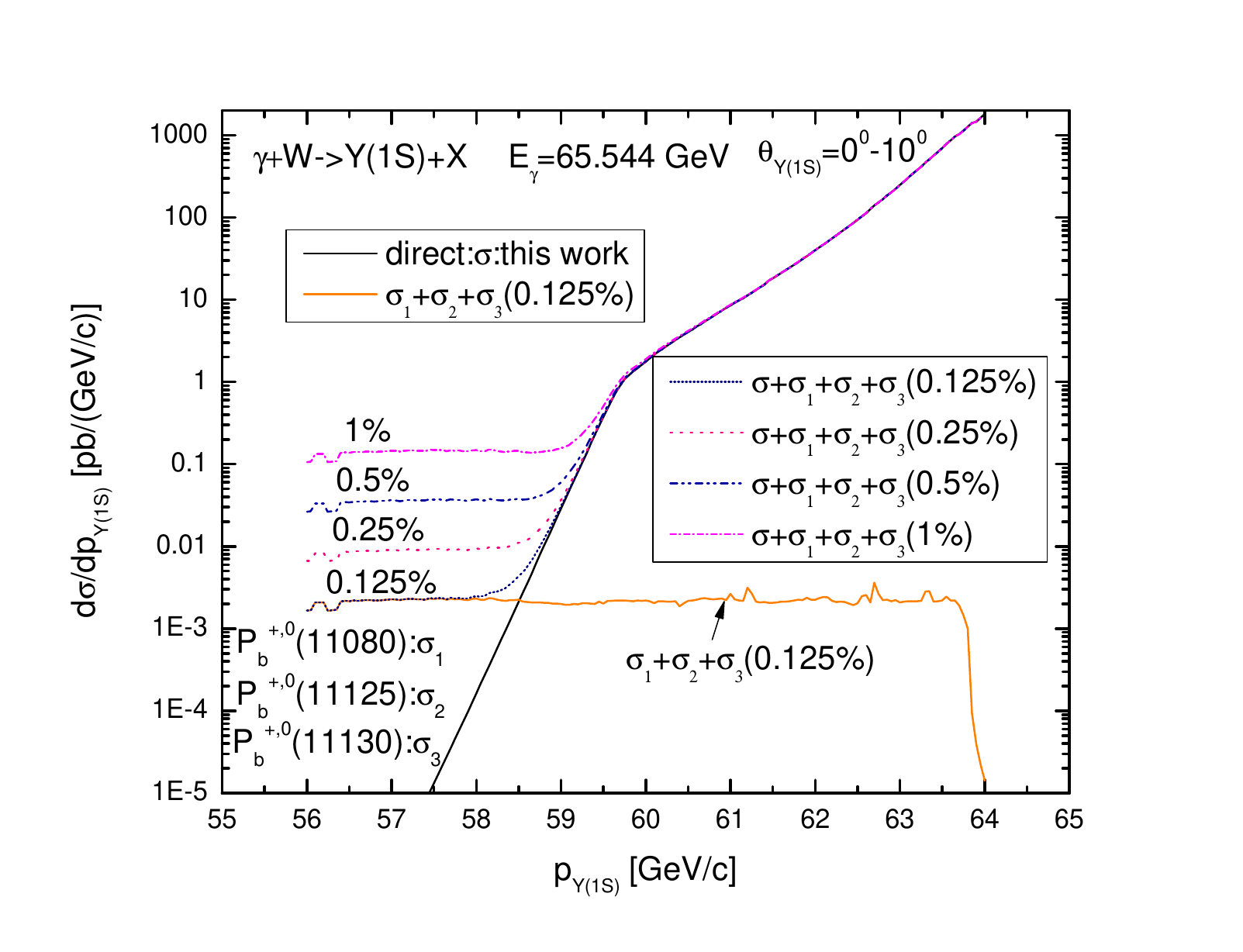}
\vspace*{-2mm} \caption{(Color online.) The same as in Fig. 11, but for the tungsten target nucleus.}
\label{void}
\end{center}
\end{figure}
%%%%%%%%%%%%%%%%%%%%%%%%%%%%%%%%%%%%%%%%%%%%%%%%%%%%%%%%%%%
\begin{figure}[h!]
\begin{center}
\includegraphics[width=16.0cm]{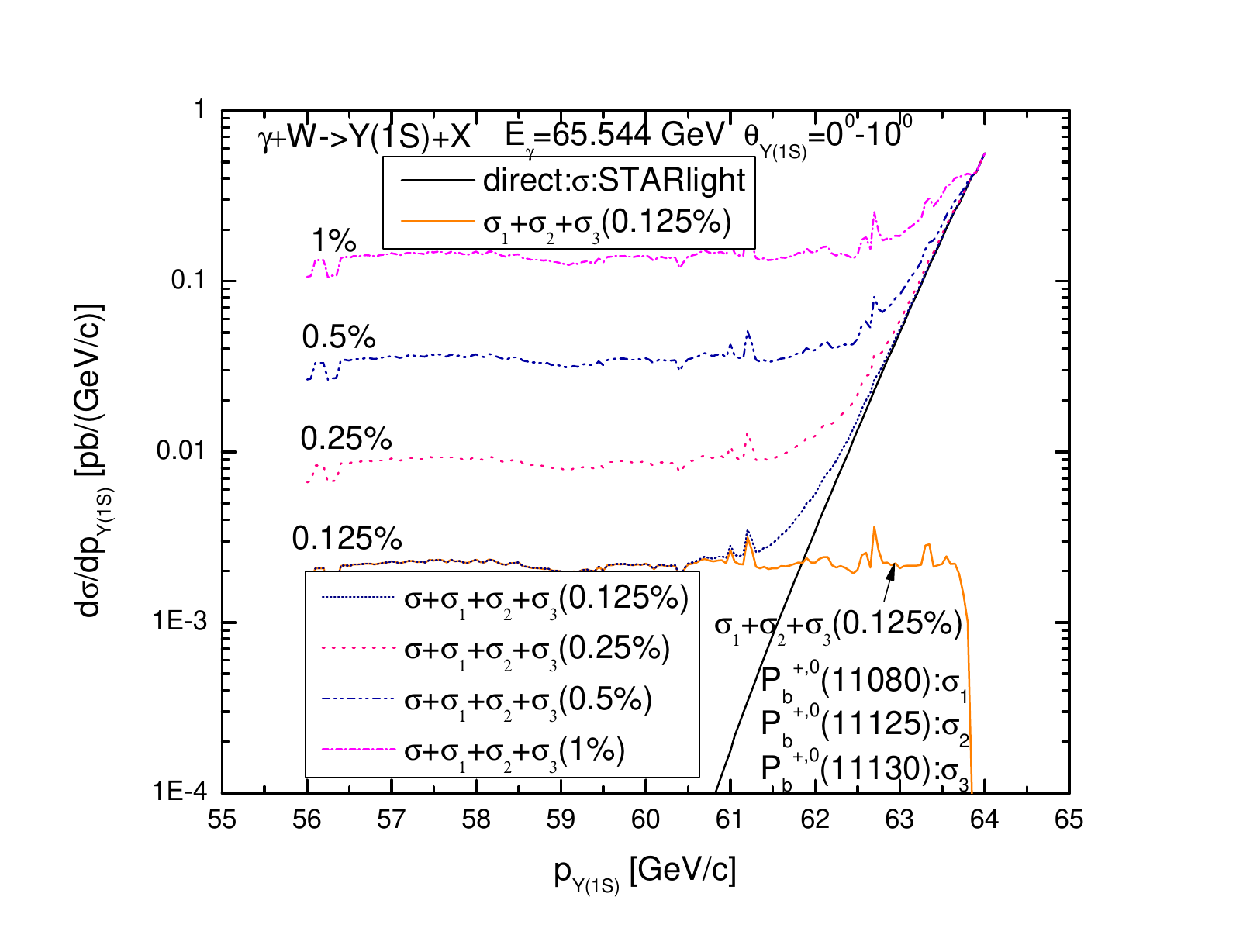}
\vspace*{-2mm} \caption{(Color online.) The same as in Fig. 12, but the non-resonant total cross section $\sigma$
for the reaction ${\gamma}p \to {\Upsilon(1S)}p$ is calculated on the basis of Eq. (45).}
\label{void}
\end{center}
\end{figure}
%%%%%%%%%%%%%%%%%%%%%%%%%%%%%%%%%%%%%%%%%%%%%%%%%%%%%%%%%%%

Finally, the absolute non-resonant, resonant and combined $\Upsilon(1S)$ meson momentum distributions,
respectively, from the direct (1), (2), two-step (53)/(55), (54)/(56) and direct plus two-step $\Upsilon(1S)$
production processes in $\gamma$$^{12}$C and $\gamma$$^{184}$W interactions, calculated on the basis of
Eqs. (50), (64) for laboratory polar angles of 0$^{\circ}$--10$^{\circ}$, for incident photon highest resonant
energy of 65.544 GeV and for the case when the total cross section of the direct process (1) is taken in the form
of Eq. (40), are shown, respectively, in Figs. 11 and 12.
The same as in Fig. 12, but determined for the total cross section of the direct process (1)
taken in the form of Eq. (45), is presented in Fig. 13 to see the sensitivity of the combined $\Upsilon(1S)$ momentum distribution to the choice of the background contribution. The resonant momentum distributions of the $\Upsilon(1S)$ mesons in the two-step processes
${\gamma}p \to P^+_{bi} \to \Upsilon(1S)p$ and ${\gamma}n \to P^0_{bi} \to \Upsilon(1S)n$ ($i=$1,2, 3),
taking place on the bound nucleons of carbon and tungsten nuclei, were obtained
for four adopted values of the branching ratios $Br[P^+_{bi} \to {\Upsilon(1S)}p]$ and
$Br[P^0_{bi} \to {\Upsilon(1S)}n]$.
Figs. 11--13 show that the total contribution to the $\Upsilon(1S)$ production on both these nuclei,
coming from the intermediate $P^+_{bi}$ and $P^0_{bi}$ states, decaying to the ${\Upsilon(1S)}p$ and
${\Upsilon(1S)}n$ channels with branching fractions of 0.125\% (orange solid curves), shows practically flat behavior,
and it is substantially larger than that from the background processes (1), (2) (black solid curves)
at momenta below approximately 58.5 and 62.0 GeV/c when the total cross section of the direct process (1) is taken in the forms of (40) and (45), respectively. As a result, at these momenta the combined $\Upsilon(1S)$ yield is completely governed by the presence of the $P^+_{bi}$ and $P^0_{bi}$ states in its production.
Its strength is almost completely determined by the branching ratios $Br[P^+_{bi} \to {\Upsilon(1S)}p]$ and
$Br[P^0_{bi} \to {\Upsilon(1S)}n]$ used in the calculations with a value, which is still large enough to be measured
at the electron-ion colliders (see above-considered), and which increases by a factor of about six when going from carbon target nucleus to tungsten one. This leads to the  well separated and experimentally distinguishable
differences between all combined calculations, corresponding to the employed options for these ratios,
for both target nuclei and for both adopted options for the background contribution.
Therefore, the $\Upsilon(1S)$ meson production differential cross section measurements
on light and especially on heavy nuclear targets in the above $\Upsilon(1S)$ "low"-momentum regions
at photon energies in the resonance regions will open an opportunity to find the hidden-bottom pentaquark
states and, if they will be observed, to determine their branching
ratios to the ${\Upsilon(1S)}p$ and ${\Upsilon(1S)}n$ modes -- at least to distinguish between realistic options of 0.125, 0.25, 0.5 and 1\%.

  In the end, in view of the above, we conclude that the near-threshold bottomonium
energy and momentum distribution measurements in photon-induced reactions both on protons
and on nuclear targets would provide evidence for the existence of the hidden-bottom pentaquarks
$P^+_{bi}$ ($P^0_{bi}$) and would give valuable information on their decay rates to the channels
${\Upsilon(1S)}p$ (${\Upsilon(1S)}n$). The similar measurements with initial electrons are also
promising at the future EicC and EIC machines. All our findings can encourage experimental groups to conduct
here such measurements.

\section*{4. Conclusions}

In the present paper we studied the near-threshold $\Upsilon(1S)$ meson photoproduction from protons and nuclei
by considering incoherent direct non-resonant (${\gamma}p \to {\Upsilon(1S)}p$,
${\gamma}n \to {\Upsilon(1S)}n$) and two-step resonant
(${\gamma}p \to P^+_{bi} \to {\Upsilon(1S)}p$, ${\gamma}n \to P^0_{bi} \to {\Upsilon(1S)}n$,
$i=1$, 2, 3; $P^{+,0}_{b1}=P^{+,0}_{b}(11080)$,
$P^{+,0}_{b2}=P^{+,0}_{b}(11125)$, $P^{+,0}_{b3}=P^{+,0}_{b}(11130)$) bottomonium production processes
with the main goal of clarifying the possibility of observation the non-strange hidden-bottom pentaquark
states $P_{bi}^{+,0}$ in this production via differential observables.
We have calculated the absolute excitation functions, energy and momentum distributions
for the non-resonant, resonant and for the combined (non-resonant plus resonant) production
of $\Upsilon(1S)$ mesons on protons as well as
on carbon and tungsten target nuclei at near-threshold incident photon energies by supposing
the spin-parity assignments of the hypothetical hidden-bottom resonances
$P^{+,0}_{b}(11080)$, $P^{+,0}_{b}(11125)$ and $P^{+,0}_{b}(11130)$ as $J^P=(1/2)^-$, $J^P=(1/2)^-$
and $J^P=(3/2)^-$ within four different realistic choices for the branching ratios
of their decays to the ${\Upsilon}(1S)p$ and ${\Upsilon}(1S)n$ modes (0.125, 0.25, 0.5 and 1\%)
as well as for two options for the background contribution. We show that these combined both integral and differential observables reveal distinct sensitivity to these scenarios and have a measurable strengths in them.
Hence, they may be an important tool to provide evidence for the existence of the above hidden-bottom pentaquark resonances as well as to clarify their decay rates. The measurements of these observables could be performed in the future at the planned electron-ion colliders EIC and EicC in the US and China. The present findings can encourage,
as one may hope, experimental groups to conduct here such measurements.

%%%%%%%%%%%%%%%%%%%%%%%%%%%%%%%%%%%%%%%%%%%%%%%%%%%%%%%%%%%%%%%%
\end{document}